\documentclass[toc]{PoS}
\usepackage{verbatim}
\usepackage{graphics, color}
\usepackage{graphicx}
\usepackage{amsmath}
\usepackage{amssymb}
\usepackage{amsthm}
\usepackage{amsfonts}
\usepackage[utf8]{inputenc}

\newcommand{\be}{\begin{equation}}
\newcommand{\ee}{\end{equation}}
\newcommand{\bfig}{\begin{figure}\begin{center}}
\newcommand{\efig}{\end{center}\end{figure}}
\newcommand{\bi}{\begin{itemize}}
\newcommand{\ei}{\end{itemize}}
\newcommand{\tr}{\mathrm{tr}}
\newcommand{\lan}{\langle}
\newcommand{\ran}{\rangle}
\newcommand{\Tr}{\mathrm{Tr}}
\newcommand{\mO}{\mathcal{O}}

\newcommand{\wt}{\widetilde}

\newcommand{\Ll}{\mathcal{L}}
\newcommand{\LR}{\mathcal{L}_R}
\newcommand{\Hh}{\mathcal{H}}

\newcommand{\Hc}{\mathcal{H}_{code}}

\newcommand{\ol}{\overline}

\newtheorem{mydef}{Definition}[section]
\newtheorem{thm}{Theorem}[section]

\newtheorem{claim}{Claim}[section]
\title{TASI Lectures on the Emergence of Bulk Physics in AdS/CFT}

\ShortTitle{The Emergence of Bulk Physics in AdS/CFT}

\author{\speaker{Daniel Harlow}\\%
       Center for Theoretical Physics, Massachusetts Institute of Technology, Cambridge, MA, 02139\\
			 Center for the Fundamental Laws of Nature, Harvard University, Cambridge, MA, 02138\\			
       E-mail: \email{harlow@mit.edu}}

\abstract{These lectures review recent developments in our understanding of the emergence of local bulk physics in AdS/CFT. The primary topics are sufficient conditions for a conformal field theory to have a semiclassical dual, bulk reconstruction, the quantum error correction interpretation of the correspondence, tensor network models of holography, and the quantum Ryu-Takayanagi formula.}

\dedicated{Dedicated to the Memory of Joe Polchinski}

\FullConference{Theoretical Advanced Study Institute Summer School 2017 "Physics at the Fundamental Frontier"\\
                 4 June - 1 July 2017\\
                 Boulder, Colorado}
\begin{document}

\tableofcontents
\section{Introduction}
The goal of these lectures is to review recent improvements in our understanding of AdS/CFT as a quantum theory of gravity.  We will be especially interested in the following questions:
\bi
\item How does local bulk effective field theory emerge from the CFT description of the theory?
\item Which CFTs does it emerge from?
\item What are the limits on its emergence?
\ei
We will focus on understanding these questions at fixed time in Lorentzian signature, using the usual quantum mechanical machinery of states and operators. The main technical results we will aim for are:
\bi
\item A set of conditions which together are sufficient for a conformal field theory to have a semiclassical dual.
\item Perturbative reconstruction of bulk fields as operators in the boundary CFT.
\item The recent reformulation of AdS/CFT in terms of error-correcting codes, and its illustration in exactly soluble toy models.  
\item The quantum extension of the Ryu-Takayanagi formula, its relation to the emergence of local bulk physics, and its information-theoretic interpretation.  Our main target here will be understanding the logic which leads to the recent proof of the ``entanglement wedge reconstruction theorem'', which gives strong evidence that AdS/CFT should be able to describe physics behind black hole horizons.
\ei
Since these ideas use many aspects of AdS/CFT, I will present the correspondence more or less from the beginning.  The approach is somewhat nonstandard, so hopefully experts might also learn something. There are many important references for this material, including \cite{Banks:1998dd,Hamilton:2006az,Heemskerk:2009pn,Heemskerk:2012mn,Heemskerk:2012np,Wall:2012uf,Papadodimas:2012aq,Faulkner:2013ana,Headrick:2014cta,Almheiri:2014lwa,Pastawski:2015qua,Jafferis:2015del,Hayden:2016cfa,Dong:2016eik,Harlow:2016vwg,Cotler:2017erl}.  In particular for following these lectures I would recommend studying \cite{Banks:1998dd,Hamilton:2006az,Heemskerk:2009pn,Heemskerk:2012mn,Almheiri:2014lwa,Harlow:2016vwg}.  Also there will be some overlap with the AdS/CFT review presented in section 6 of \cite{Harlow:2014yka}, as well as another recent review \cite{DeJonckheere:2017qkk}.  

Section \ref{adssec} reviews the basic structure AdS/CFT, section \ref{recsec} introduces bulk reconstruction, section \ref{qecsec} describes the quantum error correction intepretation of holography and gives a few toy examples, and section \ref{rtsec} introduces the quantum Ryu-Takayanagi formula and explains its relationship to quantum error correction. I close with a conclusion assessing the current status of the field.

\section{AdS/CFT Basics}\label{adssec}
The basic statement of AdS/CFT is that any conformal field theory in $d$-dimensional spacetime is equivalent to a quantum theory of gravity in a family of spacetimes which are asymptotically $AdS_d \times M$, where $M$ is some compact manifold \cite{Maldacena:1997re,Gubser:1998bc,Witten:1998qj}.  I'll begin by studying each side separately.

\subsection{Anti de Sitter space}
Anti de Sitter space (AdS) is the maximally-symmetric spacetime of constant negative curvature.  We can obtain $d+1$ dimensional AdS, denoted $AdS_{d+1}$, from a submanifold of $d+2$ dimensional Minkowski space in $(2,d)$ signature.  That space has metric
\be
ds^2=-dT_1^2-dT_2^2+dX_1^2+\ldots +dX_{d}^2,
\ee
and the submanifold we'll choose is given by
\be\label{embedding}
T_1^2+T_2^2-\vec{X}^2=\ell^2.
\ee
Here $\ell$ is parameter with units of length.  A convenient set of coordinates, called \textit{global coordinates}, are defined by
\begin{align}\nonumber
T_1&= \sqrt{\ell^2+r^2}\cos (t/\ell)\\\label{globalcoord}
T_2&= \sqrt{\ell^2+r^2}\sin (t/\ell)\\\nonumber
\vec{X}^2&=r^2,
\end{align}
which lead to the anti de Sitter space metric in global coordinates:
\be\label{adsmetric}
ds^2=-\left(1+\frac{r^2}{\ell^2}\right)dt^2+\frac{dr^2}{1+\frac{r^2}{\ell^2}}+r^2 d\Omega_{d-1}^2.
\ee
Here $r\in[0,\infty)$, $t\in (-\infty,\infty)$, and $\Omega_{d-1}$ denotes the round metric on $\mathbb{S}^{d-1}$.  

Anti de Sitter space has several important properties.  First of all it is a solution of Einstein's equation 
\be\label{EE}
R_{\mu\nu}-\frac{1}{2}R g_{\mu\nu}=8\pi G T_{\mu\nu},
\ee
with the energy-momentum tensor given by $T_{\mu\nu}=-\rho_{0}g_{\mu\nu}$, with 
\be
\rho_0=-\frac{d(d-1)}{16\pi G \ell^2}.
\ee
From here on we will usually simplify equations by working in units where $\ell=1$.  

Secondly it has a large isometry group, given by the Lorentz group of our $d+2$ dimensional Minkowski space, $SO(d,2)$.  This symmetry group acts transitively on $AdS_{d+1}$, meaning that any two points are connected by a symmetry transformation, so it is a homogeneous space.  It is also locally isotropic, meaning that the subgroup which fixes any particular point $p\in AdS_{d+1}$ can transform any tangent vector at $p$ to one which is a multiple of any other tangent vector at $p$.

\bfig
\includegraphics[height=4cm]{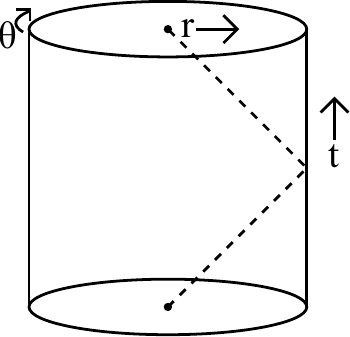}
\caption{The causal structure of $AdS_3$: a photon from the center goes out to the boundary and bounces back in time $\Delta t=\pi$.}\label{adsfig}
\efig
We can clarify the causal structure of $AdS_{d+1}$ by defining a new coordinate $\rho\in(0,\pi/2)$, via
\be
r=\tan \rho,
\ee
in terms of which we have
\be
ds^2=\frac{1}{\cos^2\rho}\left(-dt^2+d\rho^2+\sin^2\rho d\Omega_{d-1}^2\right).
\ee
Since null geodesics do not care about the Weyl factor of the metric, ie multiplication by scalar function, we see that the causal structure of $AdS_{d+1}$ will be the same as that of a ``solid cylinder'' with metric
\be
ds^2=-dt^2+d\rho^2+\sin^2\rho d\Omega_{d-1}^2.
\ee
This is illustrated in figure \ref{adsfig}.  Perhaps the most important feature of AdS space is its \textit{asymptotic boundary} at $r=\infty$ (or at $\rho=\frac{\pi}{2}$), which has topology $\mathbb{R}\times \mathbb{S}^{d-1}$.  As shown in figure \ref{adsfig}, signals can be sent to this boundary and ``replies'' received in a finite proper time according to a massive observer sitting at rest in the center of the space.  This suggests that despite the infinite spatial volume of AdS, we might want to think of it as being something like a finite box.   

If we are interested in quantum gravity, we clearly need to discuss more than just AdS space: the presence of any nontrivial matter will cause the metric to deviate from \eqref{adsmetric}.  It is natural to define a more general set of \textit{asymptotically-AdS spacetimes}, which possess an asymptotic boundary with topology $\mathbb{R}\times \mathbb{S}^{d-1}$, and whose metric approaches \eqref{adsmetric} as we approach that boundary.\footnote{For more details on how fast we need to approach it, see \cite{Henneaux:1985tv}.}  More generally it is sometimes interesting to study spacetimes which near the asymptotic boundary approach $AdS_{d+1}\times M$, where $M$ is some compact manifold whose volume stays finite (in units of $\ell=1$) as we take $r\to\infty$.

One very important example of an asymptotically-AdS spacetime is the AdS-Schwarzschild geometry, with metric
\begin{align}\nonumber
ds^2&=-f(r)dt^2+\frac{dr^2}{f(r)}+r^2 d\Omega_{d-1}^2\\
f(r)&=1+r^2-\frac{16\pi G M}{(d-1)\Omega_{d-1}}\frac{1}{r^{d-2}}.\label{adssch}
\end{align}
For $d>2$, which is where this geometry makes sense, it gives the unique family of spherically symmetric solutions of Einstein's equation \eqref{EE}.  It clearly approaches the AdS metric at large $r$, while at smaller $r$ it does something more interesting.  In fact it is so interesting that the full geometry describes not one but \textit{two} asymptotically AdS boundaries, connected by a wormhole!  More realistic geometries can be obtained by collapsing matter to form a star: the geometry is described by \eqref{adssch} outside of the star, while inside it closes off in the standard way at $r=0$.\footnote{For much more on the (AdS-)Schwarzschild geometry see, eg, \cite{Harlow:2014yka}.}

To get more intuition for AdS physics, it is very useful to study the quantization of a free scalar field in $AdS_{d+1}$, with action
\be
S=-\frac{1}{2}\int d^{d+1}x \sqrt{-g}\left(\partial_\mu \phi \partial_\nu \phi g^{\mu\nu}+m^2 \phi^2\right).
\ee
$\phi$ obeys the equation of motion
\be
\nabla^2\phi=\frac{1}{\sqrt{-g}}\partial_\mu\left(\sqrt{-g}g^{\mu\nu}\partial_\nu \phi\right)=m^2\phi.
\ee
In global coordinates it is natural to choose a basis of solutions of the form
\be
f_{\omega \ell \vec{m}}(r,t,\Omega)=\psi_{\omega \ell}(r)e^{-i\omega t} Y_{\ell \vec{m}}(\Omega),
\ee
where $Y_{\ell \vec{m}}$ are spherical harmonics on which the round sphere Laplacian acts as $-\ell(\ell+d-2)$.  $\psi_{\omega\ell}$ obeys the equation
\be\label{psieq}
(1+r^2)\psi''+\left(\frac{d-1}{r}(1+r^2)+2r\right)\psi'+\left(\frac{\omega^2}{1+r^2}-\frac{\ell(\ell+d-2)}{r^2}-m^2\right)\psi=0.
\ee
At small $r$ this equation becomes
\be
\psi''+\frac{d-1}{r}\psi'-\frac{\ell(\ell+d-2)}{r^2}\psi=0,
\ee
with solutions proportional to $r^{-\frac{d-2}{2}\pm\frac{1}{2}\sqrt{(d-2)^2+4\ell(\ell+d-2)}}$, while at large $r$ it becomes
\be
r^2 \psi''+(d+1)r\psi'-m^2 \psi=0,
\ee
which has solutions proportional to $r^{-\left(\frac{d}{2}\pm \frac{1}{2}\sqrt{d^2+4m^2}\right)}$.  Smoothness at $r=0$ requires us to choose the positive sign for the solution near $r=0$.  The sign near $r=\infty$ is more interesting: it is specified by whatever boundary conditions we choose to impose at the AdS boundary.  For $m^2\geq 0$ and $d\geq 2$ the only choice which is consistent with unitarity and preserves the full $SO(d,2)$ symmetry is to choose the plus sign:\footnote{In the limiting case $m^2=0$ and $d=2$ the minus sign is ``marginally allowed'', but it actually just describes a free \textit{boundary} scalar field, so there are no interesting bulk degrees of freedom \cite{Freivogel:2016zsb}.}
\be
\psi_{\omega \ell}(r)\sim r^{-\Delta},
\ee
with 
\be
\Delta\equiv \frac{d}{2}+\frac{1}{2}\sqrt{d^2+4m^2}.
\ee
This is called the \textit{standard quantization} of a scalar in $AdS_{d+1}$.\footnote{$m^2<0$ is sometimes also allowed by unitarity, and then the minus sign also sometimes make sense: this is called the \textit{alternate quantization} \cite{Breitenlohner:1982jf,Klebanov:1999tb}. For simplicity we will always choose the standard quantization in these lectures.}  Since we are thus imposing a nontrivial restriction on the solution of \eqref{psieq} at both $r=0$ and $r=\infty$, a quantization of $\omega$ must be required.  By solving the full equation using a hypergeometric function, we can confirm that this is indeed the case, and that in fact we must have
\be\label{omegaeq}
\omega=\omega_{n\ell}\equiv \Delta+\ell+2n, 
\ee
with $n=0,1,2,\ldots$.  

We can then write down the Heisenberg solution of our scalar field theory:
\be\label{freef}
\phi(r,t,\Omega)=\sum_{n,\ell,\vec{m}}\left(f_{\omega_{n\ell}\ell \vec{m}}a_{n\ell \vec{m}}+f_{\omega_{n\ell}\ell \vec{m}}^*a_{n\ell \vec{m}}^\dagger\right),
\ee
where $a_{n\ell\vec{m}}$, $a_{n\ell\vec{m}}^\dagger$ are annihilation/creation operators obeying
\be\label{aalg}
[a_{n\ell\vec{m}},a_{n'\ell'\vec{m}'}^\dagger]=\delta_{nn'}\delta_{\ell\ell'} \delta_{mm'}.
\ee
\eqref{aalg} is equivalent to requiring the solutions $f_{\omega_{n\ell}\ell \vec{m}}$ to have unit norm in the Klein-Gordon inner product\footnote{Here $\Sigma$ is any Cauchy slice, for example $t=0$, $n_\mu$ is the unit normal to that surface (note that $n^\mu$ is past-pointing), and $\gamma$ is the induced metric on $\Sigma$.  Also note that normalizability in the KG norm is the precise version of the ``unitarity'' which was used in the previous discussion of the boundary conditions at $r=\infty$.}
\be\label{KGn}
\lan g,f\ran\equiv i \int_{\Sigma}d^d x \sqrt{\gamma}n^\mu \left(g^*\partial_\mu f-\partial_\mu g^* f\right).
\ee
The ground state of this theory is annihilated by all of the $a_{n\ell\vec{m}}$, and ``particle'' excitations are created by acting on it with the $a_{n\ell\vec{m}}^\dagger$.  The quantization of $\omega$ tells us that, unlike in Minkowski space, the particle spectrum is discrete.  This is further evidence that we should think of physics in AdS as being analogous to living in a finite box.  

\subsection{Conformal Field Theory}
Conformal field theories (CFTs) are relativistic quantum field theories which in addition to Poincare symmetry are also invariant under \textit{dilations}
\be
x^{\mu}\,'=\lambda x^\mu
\ee
and \textit{special conformal transformations}
\be
x^{\mu}\,'=\frac{x^\mu+a^\mu x^2}{1+2a_\nu x^\nu+a^2 x^2}.
\ee
These have generators $D$ and $K_\mu$ respectively, which together with the Poincare generators $P_\mu$ and $J_{\mu\nu}$ generate the \textit{conformal group} $SO(d,2)$.  

In conformal field theories we are especially interested in \textit{primary operators}, which are local operators transforming as
\begin{align}\nonumber
e^{i D \alpha}\mO(x) e^{-i D\alpha}&=e^{\alpha \Delta}\mO\left(e^{\alpha}x\right)\\
e^{i K_\mu a^\mu} \mO(0)e^{-iK_\mu a^\mu}&=\mO(0).
\end{align}
The quantity $\Delta$ is called the \textit{scaling dimension} of $\mO$.  By acting on a primary operator $\mO$ repeatedly with derivatives we can obtain its \textit{descendant operators}, whose scaling dimensions are given by $\Delta$ plus the number of derivatives.  Descendants are never themselves primary unless they vanish.   

\bfig
\includegraphics[height=3cm]{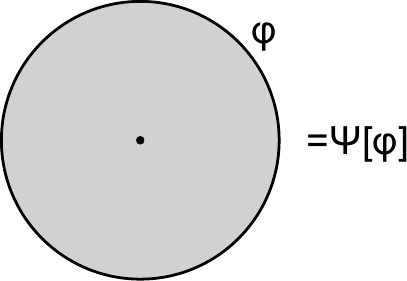}
\caption{The state-operator correspondence.  Doing path integral over the ball $\rho<1$ with boundary condition $\phi$ and an operator $\mathcal{O}$ at $\rho=0$ defines a wave functional $\Psi[\phi]$, which has sphere energy $\Delta+E_0$ if $\mathcal{O}$ has dimension $\Delta$.  Moreover given such a state, we can construct an operator that produces it by evolving the state radially inwards assuming no operators are present until we are left with something at the center that must be local.}\label{stateop}
\efig
A fundamental property of CFTs is that the set of primary operators and their descendants at any point $x$ are in one-to-one correspondence with a complete basis of the Hilbert space of the CFT quantized on $\mathbb{S}^{d-1}$: this is called the \textit{state-operator correspondence}.  The map from operators to states is defined by doing the path integral on a Euclidean solid ball centered on the operator, and it is invertible since given a state on the boundary of the ball we can do a dilatation to shrink the ball down to a point, defining an operator which would produce that state once we scale back up. Moreover if the operator we are interested in has dimension $\Delta$, then the state on $\mathbb{S}^{d-1}$ will have energy $\Delta+E_0$.  Here $E_0$ is the ground state energy: depending on the dimension we might be able to use a local counterterm to set $E_0=0$, but we also might not.  This relation is derived by observing that if we choose polar coordinates in the vicinity of the Euclidean origin and then transform $\rho=e^\tau$, we have
\be
ds^2=d\rho^2+\rho^2 d\Omega_{d-1}^2=e^{2\tau}\left(d\tau^2+d\Omega_{d-1}^2\right),
\ee
so dilations $\rho'=e^{\alpha} \rho$ are equivalent to Euclidean cylinder time translations $\tau'=\tau+\alpha$.  

\subsection{Statement of the Correspondence}
We are now ready to discuss AdS/CFT:
\begin{claim}
Any conformal field theory on $\mathbb{R}\times\mathbb{S}^{d-1}$ is equivalent to a theory of quantum gravity in asymptotically $AdS_{d+1}\times M$ spacetime, with $M$ some (possibly trivial) compact manifold.  
\end{claim}
This immediately suggests two questions:
\bi
\item[(1)] What is the map between the observables on the two sides?
\item[(2)] Which CFTs define ``semiclassical'' gravity theories, where the Planck length $\ell_P$ is much smaller than the AdS scale $\ell$ and the gravity is approximately described by Einstein's equations?
\ei
The answer to question (1) is called the \textit{dictionary}: as we will see it is still a work in progress, but many entries are known.  The most basic way to view the duality is as an isomorphism between the Hilbert spaces:
\be
\phi:\mathcal{H}_{AdS}\to \mathcal{H}_{CFT}.  
\ee
We have already seen that the asymptotic symmetry group of $AdS_{d+1}$ and the conformal group in $d$ spacetime dimensions are both isomorphic to $SO(d,2)$.\footnote{This statement on the gravity side is more subtle than it may appear: most asymptotically-AdS geometries are not invariant under $SO(d,2)$.  The right statement is that there is a set of ``large'' diffeomorphisms in the bulk which preserve the asymptotically AdS boundary conditions but act nontrivially on the asymptotic geometry and matter fields: this set is not unique, since we should quotient by ``small'' diffeomorphisms that act trivially on the asymptotic fields \cite{Henneaux:1985tv}.  The symmetry generators defined this way are local integrals at the boundary of space.}  Our first line in the AdS/CFT dictionary is that the unitary operators which implement these symmetries on the respective Hilbert spaces are related by $\phi$:
\be
\phi \circ U_{AdS}=U_{CFT}\circ \phi.
\ee
This means for example that the spectrum of the Hamiltonian is the same on both sides, and the states decompose into the same irreducible representations of $SO(d,2)$.  More generally any operator on one side can be mapped to an operator on the other using $\phi$.  To simplify equations, it is convenient to choose a basis where $\phi$ is the identity map: from now on I will do this, and I will not distinguish between states/operators written on one side or the other.  For example I will say that the Hamiltonian of the asymptotically-AdS theory is equal to the Hamiltonian of the CFT, and also that their ground states are equal.

It is difficult to formulate more pieces of the dictionary without first answering question (2): if the bulk is not semiclassical, we don't know what other observables to study.  So let's give a definition:\footnote{In this definition and the following one, for simplicity we assume that the compact extra manifold $M$ in the asymptotically-AdS boundary conditions is trivial.  We comment on how to relax this in the following subsection.}
\begin{mydef}\label{scvac}
A $d$ dimensional conformal field theory has a \textbf{semiclassical dual near the vacuum} if there exists a finite set of $CFT$ primary operators $\mO_i$ and a local bulk effective action $S_{eff}[\phi_i,\Lambda]$, with $\Lambda$ a cutoff that is large compared to $1/\ell$, but is at most $1/\ell_p\equiv G^{-\frac{1}{d-1}}$ (for both of these to be possible we need $\ell/\ell_p\gg 1$),  and $\phi_i$ are a finite set of bulk fields including the metric $g$ such that
\be\label{dictcorr}
\int \mathcal{D}\phi_i e^{iS_{eff}[\phi_i,\Lambda]}\mO_{i_1}(t_1,\Omega_1)\ldots \mO_{i_n}(t_n,\Omega_n)\approx \lan \mO_{i_1}(t_1,\Omega_1)\ldots \mO_{i_n}(t_n,\Omega_n)\ran_{CFT}
\ee 
to all orders in $\frac{1}{\ell\Lambda}$ provided that $n$ is $O(1)$ in this parameter.  Here the $\mO_i$ on the left hand side are given by the ``extrapolate dictionary''
\be\label{extd}
\lim_{r\to\infty}r^{\Delta}\phi_i(r,t,\Omega)=\mO_i(t,\Omega),
\ee
where $\Delta_i$ is the scaling dimension of $\mO_i$.  The bulk field configurations we integrate over obey the asymptotically AdS boundary conditions, with the $i\epsilon$ prescription chosen to project onto the vacuum at early and late times, and the CFT expectation values are computed in the ground state on $\mathbb{S}^{d-1}$.
\end{mydef}
This definition is sufficient to ensure that ``bulk particle physics'' arises from the CFT in an appropriate way.  The extrapolate dictionary \eqref{extd} gives a nice general relationship between bulk fields and boundary primary operators, and should be thought of as analogous to the LSZ formula in flat space. Two examples which are especially important are the boundary stress tensor $T_{\mu\nu}$, which is dual to the bulk metric $g_{\mu\nu}$, and the Noether current $J_\mu$ for any continuous boundary global symmetry, which is dual to a gauge field $A_\mu$ in the bulk.  One simple check of the reasonableness of this definition, which I encourage everyone who hasn't done it before to do, is to show the two-point function of our free scalar in $AdS$ obeys
\be
\lim_{r\to\infty}r^{2\Delta}\lan 0|\phi(r,t',\Omega')\phi(r,t,\Omega)|0\ran=\lan 0|\mO(t',\Omega')\mO(t,\Omega)|0\ran,
\ee 
where on the right hand side we have the standard CFT two point function of a scalar of dimension $\Delta$ on $\mathbb{R}\times \mathbb{S}^{d-1}$ (for help you might want to consult section 6 of \cite{Harlow:2014yka}).  

Definition \eqref{scvac} is not however quite enough to reproduce everything we know about bulk quantum gravity: it says nothing about black holes, which are excited states that are not near the vacuum and can't be produced by acting with an $O(1)$ number of $\mO_i$.  At a minimum we would like that the thermal entropy of the CFT on the sphere at sufficiently high temperature reproduces the black hole entropy we would derive from $S_{eff}$ (see eg \cite{Harlow:2014yka} for a review of these things).  A definition which is strong enough to ensure this is as follows:
\begin{mydef}\label{fulldef}
A $d$-dimensional conformal field theory has a \textbf{semiclassical dual} if in addition to having a semiclassical dual near the vacuum, \eqref{dictcorr} also holds for a more general set of asymptotically-AdS boundary conditions which allow the induced metric on the boundary to be fixed but arbitrary and possibly Euclidean: the expectation values of the $\mO_i$'s then need to match the CFT correlation functions on that boundary geometry.
\end{mydef}
In particular to get the black hole entropy to work out we need to consider the Euclidean boundary geometry $\mathbb{S}^1\times \mathbb{S}^{d-1}$:
\be
\int \mathcal{D}\phi_i e^{-S_{eff}[\phi_i,\Lambda]}\mO_{i_1}(\theta_1,\Omega_1)\ldots \mO_{i_n}(\theta_n,\Omega_n)\approx \lan \mO_{i_1}(\theta_1,\Omega_1)\ldots \mO_{i_n}(\theta_n,\Omega_n)\ran_{\mathbb{S}^1\times \mathbb{S}^{d-1}}.
\ee   
I am not aware of any example of a CFT which has a semiclassical dual near the vacuum but not more generally, so perhaps these definitions are equivalent, but in any event we will need both of them to get a full bulk picture of the CFT Hilbert space on $\mathbb{S}^{d-1}$.

\subsection{Gapped Large-N Theories}
Which CFTs have semiclassical duals?  We know a few explicit examples from string theory (up to our neglect of $M$, which I comment on at the end of this section), the most famous being the $SU(N)$ gauge theory with $\mathcal{N}=4$ supersymmetry in $d=4$, which is dual to type IIB string theory in spacetimes which are asymptotically $AdS_5\times \mathbb{S}^5$ \cite{Maldacena:1997re}.  Using the special properties of this theory implied by supersymmetry, one can give strong evidence that the appropriate generalization of \eqref{scvac} to include $M=\mathbb{S}^5$ holds.\footnote{In more detail, we can observe that this CFT has a set of ``chiral BPS operators'' whose scaling dimensions are protected by supersymmetry, and which match the fields of IIB supergravity in the bulk \cite{Witten:1998qj}.  The equivalence of correlation functions \eqref{dictcorr} can be checked explicitly in simple cases, and the general equivalence should follow from the large-$N$ analysis I'll discuss in a moment, extended to nontrivial $M$.  For the more general boundary manifolds required by definition \eqref{fulldef} less has been explicitly confirmed, but in some other examples this has also been checked.  The general qualitative behavior does seem to match \cite{Witten:1998qj,Witten:1998zw}.}  It would be nice however to have a general criterion on CFTs which is necessary and sufficient for the existence of a semiclassical dual.  We do not yet have such a criterion, but we do have a criterion that seems likely to be \textit{sufficient}:\footnote{The terminology here is a bit unfortunate, since a condensed-matter physicist would say that any CFT is ``gapless''.  Here we mean that there is a gap in the spectrum of CFT primary operators from the multitrace primaries to everything else, not a gap in the spectrum of the Hamiltonian on $\mathbb{R}^d$.  Both terminologies are standard in their respective communities, so I see no way to avoid this.}
\begin{mydef}\label{Ndef}
A \textbf{gapped large-N CFT} is a family of CFTs labeled by a parameter $N$ such that:
\bi
\item There is a finite set of ``single-trace'' primaries $\mO_i$ with the property that if we normalize them so that $\lan\mO_i \mO_i\ran\sim N^0$ then 
\be\label{N3pt}
\lan \mO_i \mO_j \mO_k\ran\lesssim \frac{1}{N^{\#/2}},
\ee
with $\#$ some $O(N^0)$ number.  
\item There is only one single-trace primary of spin $2$ and scaling dimension $d$, the energy-momentum tensor $T_{\mu\nu}$, and its three point function has a nonzero term of order $\frac{1}{N^{\#}/2}$ (this contribution could be zero for other operators).    
\item For any collection $\{\mO_{i_1},\ldots,\mO_{i_n}\}$ of single-trace operators, with $n\sim N^0$, there is an associated ``multi-trace'' primary $\mO_{i_1\ldots i_n}$ with dimension $\Delta_{i_1}+\ldots+\Delta_{i_n}$.
\item At leading order in $1/N$, correlation functions of single- and multi-trace operators can all be computed by Wick contraction.  For example:
\begin{align}\nonumber
\lan\mO_i\mO_j \mO_i \mO_j\ran&=\lan\mO_i \mO_i\ran \lan\mO_j \mO_j\ran \qquad \qquad (i\neq j)\\
\lan \mO_i \mO_j \mO_{ij}\ran&=\lan\mO_i\mO_i\ran \lan \mO_j\mO_j\ran \qquad \qquad (i\neq j).\label{Nfac}
\end{align}
Moreover the corrections to this statement are at most of order $N^{-\#(n-2)/2}$, where $n$ is the number of single-trace operators plus the number of multitrace operators counted including multiplicity (eg $n=4$ for both examples in \eqref{Nfac}).
\item All operators with $\Delta\sim N^0$ are single/multi-trace primaries and their descendants.

\ei
\end{mydef}
We can then make a claim \cite{Heemskerk:2009pn}:
\begin{claim}
Any gapped large-$N$ theory has a semiclassical dual. The bulk fields $\phi_i$ correspond to the single-trace primaries $\mO_i$. 
\end{claim}
This claim is somewhat analogous to the old lore in particle physics that every Lorentz-invariant S-matrix with a finite set of particle species and obeying cluster decomposition corresponds to a local quantum field theory \cite{Weinberg:1995mt}.  It has not  been proven, but there is quite a bit of evidence for it.  Let's first understand more about what kind of $S_{eff}$ we should expect to reproduce.  There will be finitely many fields $\phi_i$ dual to the $\mO_i$.  Comparing the stress-tensor three-point function to the three-graviton vertex tells us that we must have
\be\label{Geq}
G\ell^{-(d-1)}\sim \frac{1}{N^{\#}},
\ee
so $N$ controls the strength of gravitational interactions.  The conditions around \eqref{N3pt}, \eqref{Nfac} then tell us that all bulk interactions between the $\phi_i$'s are of at most gravitational strength.\footnote{This is a rather strong restriction, for example it says there will be no interesting bulk particle physics.  This is the main reason that this criterion is not necessary for a CFT to have a semiclassical dual: for example in the string landscape we think there are many AdS vacua with interesting particle physics that arise from more sophisticated compactifications.} The details of the current evidence that any gapped large-$N$ theory indeed has such a bulk $S_{eff}$ are beyond the scope of these lectures, but the basic idea is to use the conformal bootstrap to parameterize all solutions of the four-point function crossing equation for single- and multi-trace primaries, and then show that each parameter corresponds to the coefficient of a potential local term in $S_{eff}$ \cite{Heemskerk:2009pn}.  Some aspects of this argument are greatly simplified in Mellin space \cite{Penedones:2010ue}, but despite continued progress (see for example \cite{Aharony:2016dwx}) the full story hasn't quite been nailed down yet.  The checks are convincing enough however that for the remainder of these notes we will assume that this claim is correct.  

Before moving on, I need to briefly address my neglect of the extra compact manifold $M$ in definitions \eqref{scvac}, \eqref{Ndef}. The few explicit examples of AdS/CFT we actually know, such as the $\mathcal{N}=4$ theory with $d=4$ or the ABJM theory with $d=3$ \cite{Aharony:2008ug}, always asymptote to $AdS_{d+1}\times M$ with $M$ nontrivial.  \eqref{scvac}, \eqref{Ndef} are then modified in the following way: each bulk field then corresponds not to one primary but instead to an infinite (at large $N$) tower of primaries that distinguish where the bulk field goes on $M$ as we take $r\to\infty$.  I could have included this possibility into the definitions all along, but it would have made the following discussion more tedious so I have chosen not to.  Theories with trivial $M$ should be obtainable by compactifying string theory on Calabi-Yau manifolds with flux \cite{Klebanov:2000hb,Giddings:2001yu,Kachru:2003aw}, but so far we do not know an independent way of constructing their dual CFTs.  It would be interesting to redo the analysis of \cite{Heemskerk:2009pn,Penedones:2010ue} for a more general set of gapped large-$N$ theories which allow nontrivial $M$, but in the meantime I will proceed continuing to assume that $M$ is trivial.

Finally let's consider the high-energy structure of gapped large $N$ theories.  We have not yet said much about the interpretation of the parameter $N$ in the CFT.  Equation \eqref{Geq} suggests that since $1/G$ appears in front of the bulk action, we might expect $N^{\#}$ to count degrees of freedom in the CFT.  Following this suggestion, and using that at high temperature the partition function of the CFT should be extensive in the volume of the sphere in units of temperature, we can guess that
\begin{align}
S&\sim N^{\#}T^{d-1}\\
E&\sim N^{\#} T^d.
\end{align}
Eliminating $T$, we have
\be\label{Scft}
S\sim N^{\#/d}E^{\frac{d-1}{d}}.
\ee
We can compare this to our bulk expectations, which are that states with $T\gg 1$ should correspond to big black holes.  For these black holes we have
\be
r_s^d\sim GM,
\ee
and the Bekenstein-Hawking formula then tells us that we should expect
\be\label{Sbulk}
S=\frac{A}{4G}\sim \frac{r_s^{d-1}}{G}\sim G^{-1/d}M^{\frac{d-1}{d}}.
\ee
Using \eqref{Geq} we see that \eqref{Scft} and \eqref{Sbulk} indeed match!  This is a qualitative illustration of the expectation that quantum gravity is holographic: the real number of degrees of freedom is subextensive in the bulk volume.  This heuristic analysis can be confirmed at the level of scaling in the $\mathcal{N}=4$ theory, since Yang-Mills fields are $N\times N$ matrices, which suggests $\#=2$ in the thermodynamic quantities, while the usual planar diagram analysis confirms that indeed $\#=2$ in the stress tensor three-point function.  In some special cases we can even match the $O(1)$ coefficient, giving a microscopic derivation of the Bekenstein-Hawking entropy formula \cite{Strominger:1997eq,Hartman:2014oaa,Benini:2015eyy}.

\section{Bulk Reconstruction}\label{recsec}
\bfig
\includegraphics[height=4cm]{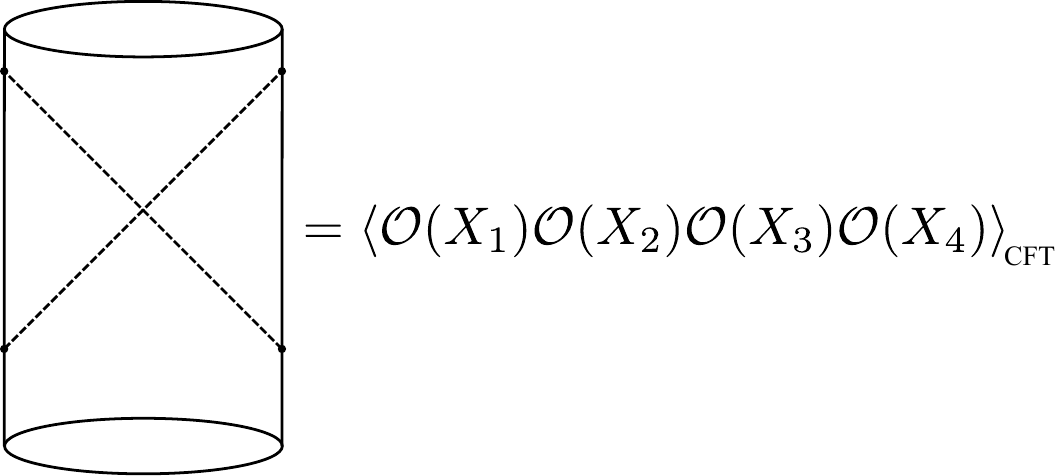}
\caption{Using the extrapolate dictionary to describe a bulk scattering experiment to the CFT.}\label{ads4pt}
\efig
The dictionary we have developed so far is sufficient for discussing ``scattering'' experiments where we act with boundary local operators at some early time, wait for a while, and then measure some boundary operators to see what comes out.\footnote{It also is sufficient for more general ``out of time order scattering experiments'', where between local measurements we allow ourselves to evolve the CFT backwards as well as forwards in time.  Recently an interesting connection between such experiments and quantum chaos has been re-discovered, leading to several new insights on the nature of chaos \cite{Maldacena:2015waa}.}  We illustrate such an experiment in figure \ref{ads4pt}.
A lot of very interesting physics is contained in such experiments, perhaps most importantly the unitarity of black hole formation and evaporation \cite{Witten:1998qj,Maldacena:2001kr} (see also \cite{Harlow:2014yka} for some review).  

\bfig
\includegraphics[height=4cm]{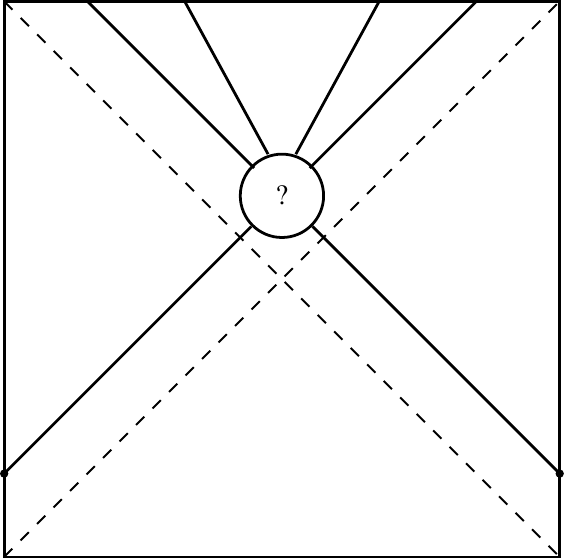}
\caption{A bulk experiment we don't know how to describe using correlation functions of local operators in the CFT.  This is the Penrose diagram of the two-sided AdS-Schwarzschild geometry, we have prepared an initial state with two incoming particles, and we would like to know the probability distribution for what they produce behind the horizon.}\label{bhscatter}
\efig
Unfortunately it does not seem to be the case that \textit{all} interesting experiments in the bulk can be so easily related to local correlation functions in the CFT.  The problem is that the output of an experiment which happens behind a horizon is by definition unable to reach the boundary, at least not if the semiclassical picture of the spacetime is correct.  A simple example is shown in figure \ref{bhscatter}.  This suggests that it would be useful to understand how to define operators in the CFT which represent bulk operators which are far from the boundary: we'd like to back off of the extrapolate dictionary \eqref{extd}.  Producing such a representation is somewhat mysteriously called \textit{bulk reconstruction}.  

There are two immediate objections you might have to the feasibility of bulk reconstruction:
\bi
\item Local bulk operators like $\phi(t,r,\Omega)$ are not diffeomorphism invariant, so why should they mean anything in the CFT, where all states are physical?
\item We know from black holes that the maximal entropy in a region scales like the area of its boundary, and indeed we confirmed this in the CFT in the previous section under plausible assumptions.  So doesn't this mean that we shouldn't really be able to describe the bulk using a local Lagrangian, and shouldn't local bulk operators not really be well-defined?
\ei
The first of these problems is not so serious, roughly speaking we can deal with it by gauge-fixing.  More carefully we need to define bulk operators including a ``gravitational dressing'' which attaches them to the boundary, analogous to in electrodynamics where a charged operator must be attached to infinity by a Wilson line.  I will discuss this occasionally as we go on, but for the most part I will leave this dressing implicit.  The second problem is a deeper one: understanding some of its implications is the main goal of these lectures.  Ultimately we will see that our bulk operators are defined only up to non-perturbative ambiguities that are at least of order $e^{-1/G}$.  Moreover we will see that in general they should only be understood as acting on a subspace of states in the CFT where we expect them to make sense in the bulk.  

\subsection{Free Scalar Reconstruction}
Our initial approach to reconstruction will be based on the following simple proposal: reconstruct bulk operators like $\phi(t,r,\Omega)$ by perturbatively solving the bulk equations of motion as operator equations in the CFT, using the extrapolate dictionary \eqref{extd} to set the boundary conditions \cite{Banks:1998dd,Hamilton:2006az,Kabat:2011rz,Kabat:2012hp,Heemskerk:2012mn,Heemskerk:2012np,Kabat:2013wga}. We'll begin with the simple case of a free scalar in $AdS_{d+1}$. 

Recall from \eqref{freef} that we can expand a free field in terms of creation and annihilation operators
\be\label{freef2}
\phi(r,t,\Omega)=\sum_{n\ell \vec{m}}\left(f_{n\ell\vec{m}}a_{n\ell\vec{m}}+f_{n\ell\vec{m}}^*a_{n\ell\vec{m}}^\dagger\right),
\ee
with the (slightly relabelled) Klein-Gordon solutions 
\be
f_{n\ell\vec{m}}(r,t,\Omega)=\psi_{n\ell}(r)e^{-i\omega_{n\ell}\,t}Y_{\ell\vec{m}}(\Omega).
\ee
The boundary conditions at $r=0$ and $r=\infty$ give the quantization $\omega_{n\ell}=\Delta+\ell+2n$ and the asymptotic behavior
\be\label{psia}
\psi_{n\ell}(r)=N_{n\ell}r^{-\Delta}\left(1+O(r^{-2})\right).
\ee
The constant $N_{n\ell}$ is determined by normalizing these modes to unity using the KG inner product \eqref{KGn}, we won't need its explicit form here but an expression can be found for example in \cite{Harlow:2014yka}.  Substituting \eqref{freef2} and \eqref{psia} into the extrapolate dictionary \eqref{extd}, we find that
\be
\mathcal{O}(t,\Omega)=\sum_{n\ell\vec{m}}\left(N_{n\ell}e^{-i\omega_{n\ell}\,t}Y_{\ell \vec{m}}(\Omega)a_{n\ell\vec{m}}+h.c.\right)
\ee
In what follows it is convenient to separate this into positive and negative frequency parts, $\mathcal{O}=\mO_++\mO_-$, with
\be
\mO_+(t,\Omega)=\sum_{n\ell\vec{m}}N_{n\ell}e^{-i\omega_{n\ell}\,t}Y_{\ell \vec{m}}(\Omega)a_{n\ell\vec{m}}.
\ee
Taking the Fourier transform, we see that
\be
a_{n\ell\vec{m}}=N_{n\ell}^{-1}\int_{-\pi}^\pi dt\, e^{i\omega_{n\ell}\,t}\int d\Omega \,Y_{\ell\vec{m}}^*(\Omega)\mO_+(t,\Omega).
\ee
We can then substitute this back into \eqref{freef2}, to find \cite{Banks:1998dd}
\be\label{phir1}
\phi(r,t,\Omega)=\int_{-\pi}^{\pi} \, dt'\int d\Omega'\, K_+(r,t,\Omega;t',\Omega')\mO_+(t',\Omega')+h.c.\,,
\ee
where
\be
K_+(r,t,\Omega;t',\Omega')\equiv\sum_{n\ell\vec{m}}N_{n\ell}^{-1}f_{n\ell\vec{m}}(r,t,\Omega)e^{i\omega_{n\ell}\,t'}Y_{\ell\vec{m}}^*(\Omega').
\ee
This gives our first (of several) CFT representations of $\phi(r,t,\Omega)$. 

The reconstruction \eqref{phir1} does give us a way to represent the bulk field $\phi$ in the CFT, but it has two somewhat strange features.  First of all there is the annoying split into positive and negative frequency parts: this arose because the modes $e^{i\omega_{n\ell} t}$ and $e^{-i\omega_{n'\ell'} t}$ are not necessarily orthogonal unless $\Delta$ is an integer, which is not usually the case, so we had to do the Fourier transform separately for positive and negative frequencies.  This would not be an issue if the function $K_+$ were real, but in general it isn't.  Secondly, it is somewhat counter-intuitive that the range of $t'$ integration is independent of $t$: we could be interested in an operator at $t=10^{50}$, and we would still just integrate $t'$ from $-\pi$ to $\pi$.  These are not inconsistencies, but it turns out that there is an ambiguity in the construction of $K_+$ which we can take advantage of to clean up both of these issues.  Indeed we are free to add to $K_+$ any function which integrates to zero against $\mO_+$: in particular any function whose frequency components are of the form $e^{i(\Delta-m)t'}$, with $m$ an integer \cite{Hamilton:2006az}.

To ``fix'' these problems, we first observe that if we choose to place our bulk operator at $r=t=0$, then the boundary region integrated over in \eqref{phir1} is just the set of all boundary points which are spacelike-separated from that bulk point.  For this point we have
\be
K_+(0;t',\Omega')=\frac{1}{\Omega_{d-1}}\sum_n N_{n0}^{-1}\psi_{n0}(0)e^{i(\Delta+2n)t'}.
\ee
Using the explicit formula for $\psi_{n\ell}$ this sum can be evaluated in terms of a hypergeometric function, and by using the freedom to shift by multiples of $e^{i(\Delta-m)t}$ we may then redefine it in such a way that $K_+$ is real \cite{Hamilton:2006az}.  The positive and negative frequency parts may now be combined, and we may then use the AdS isometry to move this bulk point to any other bulk point, generating a new smearing function $K(r,t,\Omega;t',\Omega')$ in terms of which we have
\be\label{globalR}
\phi(x)=\int_{S_x} dX K(x;X)\mO(X).
\ee
Here I have defined an abbreviated notation where $x=(r,t,\Omega)$ is a bulk point and $X=(t',\Omega')$ is a boundary point, $dX$ implicitly contains the usual  factor $\sqrt{-\gamma}$, with $\gamma$ the determinant of the boundary metric, and $S_x$ is the set of boundary points which are spacelike separated from $x$.  I will not present the details of this argument here, since we will arrive at the same conclusion on more general grounds in the following section.

\bfig
\includegraphics[height=6cm]{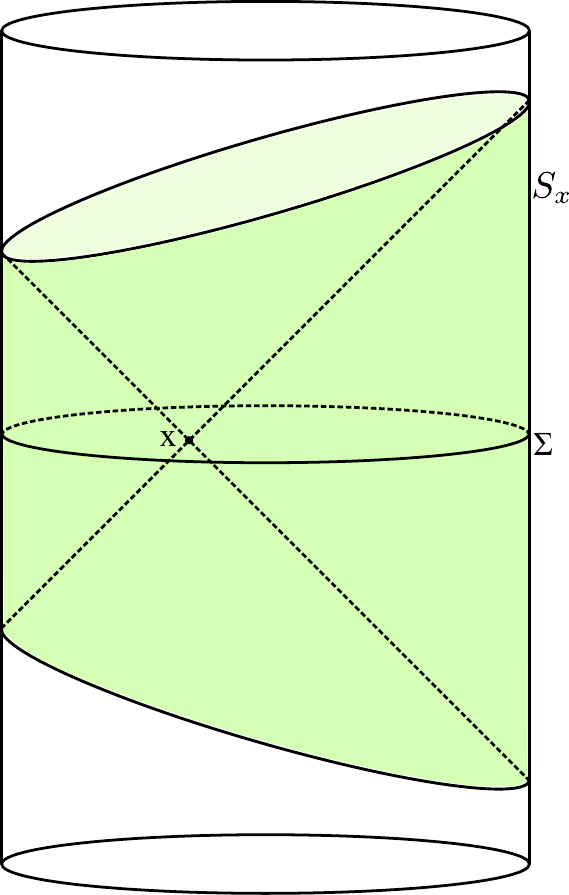}
\caption{\textbf{Global Reconstruction} A bulk scalar field $\phi(x)$ is represented in the CFT (at leading order in the interactions) as an integral of the CFT local operator $\mO$ dual to $\phi$, integrated over the set of boundary points $S_x$ that are spacelike separated from $x$, shaded here in green.  $S_x$ has nontrivial support on an entire Cauchy slice of the boundary, denoted $\Sigma$.} \label{global}
\efig
The reconstruction \eqref{globalR} has the property that it represents the bulk operator $\phi$ in a way that uses the CFT degrees of freedom on an entire Cauchy slice $\Sigma$, so it is called \textit{global reconstruction}.  We illustrate it in figure \ref{global} .  The function $K$ is sometimes called a \textit{smearing function}: it is a kernel we smear against the boundary operator $\mO$ to generate the bulk operator $\phi$.\footnote{To avoid confusion I want to emphasize that $K$ is \textit{not} the ``bulk-to-boundary propagator'' often considered in the AdS/CFT literature. We will see the difference in more detail in the following section.}  Note that the CFT representation of $\phi$ is nonlocal both in space and time: by using the CFT time evolution we could choose to represent it ``in the Schrodinger picture'' using only operators on $\Sigma$, but arriving at this expression would require us to solve the CFT time evolution.  In general this will produce operators at $\Sigma$ which are very nonlocal (not just a single integral over space).  

\subsection{Including Interactions}
There is clearly more to life than free field theory near the vacuum.  For one thing, the two-point function of a scalar field in AdS is ultimately determined by conformal invariance, so in fact the bulk scalar field we reconstructed in the previous section could have been reconstructed just as well in \textit{any} boundary conformal field theory, even one without a semiclassical dual!  The real power of the bulk reconstruction procedure based on solving equations of motion is that it can be extended to perturbatively include interactions \cite{Kabat:2011rz}; this is what sets it apart from other reconstruction proposals such as \cite{Verlinde:2015qfa,Nakayama:2015mva,Czech:2016xec,Anand:2017dav} which are based purely on using conformal (or Virasoro) symmetry.  The streamlined treatment we give here is basically verbatim taken from \cite{Heemskerk:2012mn}.  I will for simplicity discuss only an interacting scalar field, the extension to higher spins is mostly straightforward \cite{Kabat:2012hp,Kabat:2012av,Heemskerk:2012np,Kabat:2013wga}.  

The key idea is to introduce the notion of a \textit{spacelike bulk Green function}, obeying
\begin{align}\label{spG}
(\Box'-m^2)G(x,x')&=\frac{1}{\sqrt{-g}}\delta^{d+1}(x-x')\\
G(x,x')&=0 \qquad (x,x' \mathrm{\,\,not \,\,spacelike\,\,separated})
\end{align}
Here and throughout this section I will use the abbreviated notation of equation \eqref{globalR}, where $x$ denotes a bulk point, $X$ a boundary point, $dx$ the appropriate integration measure $d^{d+1} x \sqrt{-g}$ in the bulk, and $dX=d^{d}X\sqrt{-\gamma}$ the appropriate integration measure in the boundary.  The spacelike support required of this Green function is nonstandard, and it is not obvious that such a Green function exists.  In $AdS_{d+1}$ however, if we choose $r=0$ then by symmetry the Green function can depend only on $r'$ and $t'-t$.  This reduces to a $1+1$ dimensional boundary value problem, where we can exchange $r'$ and $t'-t$ to get a standard causal Green function problem.  This ensures that $G(x,x')$ exists with spacelike support provided that $r=0$, and then we may use the AdS symmetry to move this solution, obtaining a spacelike Green function for arbitrary $(x,x')$ \cite{Heemskerk:2012mn}.  Beware however that this $G(x,x')$ is not symmetric between $x$ and $x'$, and beware also that it is not equal to the usual ``bulk-to-bulk propagator'' defined by the time-ordered two-point function of a free massive scalar in AdS. That Green function \textit{is} symmetric between the two points, and it falls off as $r^{-\Delta}$ as we move either point to the boundary.  By contrast as $r'\to\infty$ our spacelike Green function obeys
\be\label{Gasymp}
G(x,x')\sim \frac{1}{2\Delta-d}\left(r'^{-\Delta} L(x,X')+r'^{-(d-\Delta)}K(x,X')\right),
\ee
with $K(x,X')$ a nonzero function which has support only when $x$ and $X'$ are spacelike separated.

\bfig
\includegraphics[height=3cm]{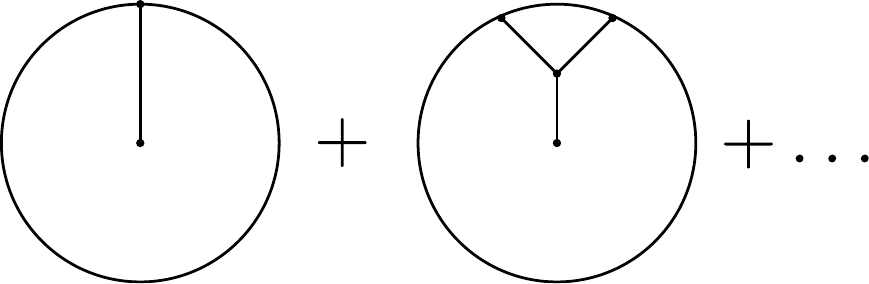}
\caption{Diagrammatic bulk reconstruction in the presence of interactions.  Boundary points are connected to bulk points using the smearing function $K$, while bulk points are connected to bulk points using the spacelike Green function $G$.}\label{higherorderfig}
\efig
Using this spacelike Green function we can use integration by parts to arrive at the following simple identity
\begin{align}\nonumber
\phi(x)&=\int dx' \phi(x')\left(\Box'-m^2\right)G(x,x')\\\nonumber
&=\int dX' r^\mu\left(\phi(x')\partial'_\mu G(x,x') -G(x,x')\partial'_\mu \phi(x')\right)+\int dx'G(x,x')\left(\Box'-m^2\right)\phi(x')\\\label{iter}
&=\int dX'K(x,X')\mO(X')+\int dx'G(x,x')\left(\Box'-m^2\right)\phi(x'),
\end{align}
where in the third equality we used the extrapolate dictionary \eqref{extd} and the asymptotic form \eqref{Gasymp}.  If $\phi$ is free then the second term vanishes, so we recover our previous reconstruction \eqref{globalR} in a more elegant manner.  More generally we can iterate equation \eqref{iter} to derive a perturbative expression for $\phi(x)$ in terms of $\mO(x)$.  For example if the bulk effective action $S_{eff}$ appearing in \eqref{dictcorr} gives $\phi$ a nonlinear equation of motion
\be
(\Box-m^2)\phi=g\phi^2,
\ee 
then iterating once we have
\begin{align}\nonumber
\phi(x)=&\int dX'K(x,X')\mO(X')+g\int dX' dX''dx' K(x',X')K(x',X'')G(x,x')\mO(X')\mO(X'')\\
&+O(g^2).\label{phi3eq}
\end{align}
This expansion has a Feynman-like graphical representation, shown in figure \ref{higherorderfig}.  At higher orders this reconstruction becomes more and more nonlocal, and soon starts bringing in other single-trace operators dual to bulk fields that $\phi$ couples too.  In particular the stress tensor $T_{\mu\nu}$ will make an appearance, since $\phi$ is always coupled at least to gravity \cite{Kabat:2013wga}.    If the CFT is a large $N$ theory as in definition \ref{Ndef}, then the coupling $g$ will scale like an inverse power of $N$.

It is interesting to consider what would go wrong if we used this procedure to construct an operator solution in the CFT of a set of equations of motion \textit{other} than the ones derived from $S_{eff}$.  We would not fail in constructing a solution, but what would fail is the algebra of the bulk operators. In particular, commutativity at spacelike separation would fail \cite{Kabat:2011rz}.  For example say that $[\phi(x),\phi(y)]$ had a nonzero contribution at spacelike separation proportional to $\phi$.  We could detect this by studying the three point function $\lan [\phi(x),\phi(y)] \phi(z)\ran$, which by the extrapolate dictionary must be proportional to the three-point function coefficient $C_{\mO\mO\mO}$ in the CFT.  In fact if we use the uncorrected reconstruction \eqref{globalR} for $\phi$ then the CFT three-point function will indeed predict such a nonzero commutator proportional to $C_{\mO\mO\mO}$: the $O(g)$ term in \eqref{phi3eq} will only cancel this commutator if we take $g\sim C_{\mO\mO\mO}$ with just the right coefficient \cite{Kabat:2011rz}.  

At sufficiently high orders in this expansion, we will eventually need to confront the gauge non-invariance of $\phi(x)$ \cite{Kabat:2012av,Heemskerk:2012np,Kabat:2013wga}.  For example if there is a $U(1)$ gauge symmetry in the bulk, under which $\phi$ transforms as
\be
\phi'(x)=e^{i\Lambda(x)}\phi(x),
\ee
then for example we can attach $\phi$ to the boundary with a Wilson line, creating a gauge-invariant ``dressed'' operator
\be\label{dressedphi}
\wt{\phi}(x)\equiv e^{i \int_C A}\phi(x),
\ee
where $C$ is some curve connecting $x$ to a point on the boundary.  A similar dressing is necessary to deal with diffemorphism invariance: the closest analogy to this Wilson line is to place the operator $\phi$ at a the endpoint of a spatial geodesic fired orthogonally into the bulk from a boundary point $X$ and with renormalized proper length $L$.  This amounts to studying operators in ``holographic gauge'', where the metric is constrained to be of the form
\be
ds^2=\frac{dr^2}{r^2+1}+g_{ij}dx^i dx^j,
\ee
with $i$, $j$, running over $t$ and $\Omega$.  

Finally I'll briefly discuss perturbative expansions around backgrounds other than pure AdS.  In principle the same construction should work, but with the notable defect that the argument for the existence of the spacelike Green function given above relied crucially on the symmetries of AdS.  The existence of spacelike Green functions for generic asymptotically AdS geometries is an unsolved problem in linear PDEs, and it will remain so at least until someone who actually knows something about general linear PDEs works on it.  In the meantime however there is one important point which can be made: this is that in principle the expansion around AdS \textit{does} actually contain all the information we need to work out the operators on other backgrounds.  An analogy to quantum electrodynamics is in order here. Consider the motion of an electron in the presence of a heavy positively-charged particle such as a proton: we are taught in childhood that we should deal with this by studying the motion of the electron in the classical Coulomb field sourced by the proton.  The electron will have a propagator in this background which is different from that in vacuum.  But this treatment must be equivalent to what we would have gotten by treating the proton as dynamical and doing conventional QED, with standard vacuum propagators.  Indeed it is, the former description is obtained by summing an infinite series of ``ladder'' diagrams in the latter description (see eg section 13.6 of \cite{Weinberg:1995mt}).  So it is here: by summing vacuum bulk reconstruction diagrams, we can in principle show that our reconstructed operator also holds in other geometries.  The details are more complicated than for electrodynamics, for a check at first nontrivial order see \cite{Duff:1973zz}.

\subsection{AdS/Rindler Reconstruction}
So far we have been discussing global reconstruction, whereby a bulk field $\phi(x)$ is given a CFT representation that has support on an entire boundary timeslice.  In this subsection we will see that sometimes an alternative representation is possible, where the spatial support of the CFT representation is restricted to some proper subregion of a boundary time slice \cite{Hamilton:2006az,Morrison:2014jha}.  This will be our first iteration of the powerful idea of \textit{subregion duality}: the notion a spatial subregion $R$ of the boundary CFT has complete information about what is going on in a yet-to-be-determined subregion of the bulk \cite{Bousso:2012sj,Czech:2012bh,Bousso:2012mh}.  Determining what this bulk subregion is will be one of the primary goals of these lectures.  

We'll begin by recalling our identification of $AdS_{d+1}$ as an embedded submanifold in $(2,d)$ signature Minkowski space via equation \eqref{embedding}.  So far we have parametrized this embedding via the ``global'' coordinates \eqref{globalcoord}, leading to the metric \eqref{adsmetric}.  But now we will instead choose a different parametrization:
\begin{align}\nonumber
T_1&=\rho \cosh \chi\\\nonumber
T_2&=\sqrt{\rho^2-1}\sinh \tau\\\nonumber
X_d&=\sqrt{\rho^2-1}\cosh \tau\\
X_1^2+\ldots X_{d-1}^2&=\rho^2\sinh^2 \chi,
\end{align}
in terms of which the metric becomes
\be
ds^2=-(\rho^2-1)d\tau^2+\frac{d\rho^2}{\rho^2-1}+\rho^2 dH_{d-1}^2,
\ee
where
\be
dH^2_{d-1}=d\chi^2+\sinh^2\chi d\Omega_{d-2}^2
\ee
is the metric on the $(d-1)$-dimensional hyperbolic ball.  In these coordinates we take $\rho>1$, and $-\infty<\tau<\infty$.  The parametrize a subregion of the full $AdS_{d+1}$, called the AdS-Rindler wedge, which is shaded blue in figure \ref{rindlerfig}.
\bfig
\includegraphics[height=5cm]{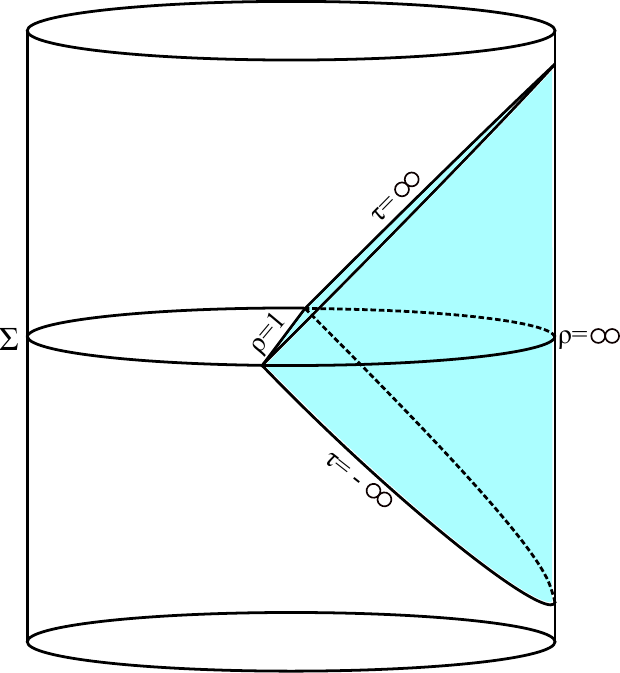}
\caption{The AdS-Rindler wedge.}\label{rindlerfig}
\efig

As in our global description of $AdS_{d+1}$, we can quantize a scalar field in the Rindler wedge.  We do this by expanding in modes, via
\be
\phi(\rho,\tau,\alpha)=\int_0^\infty \frac{d\omega}{2\pi}\sum_\lambda\left(f_{\omega\lambda}(\rho,\tau,\alpha)a_{\omega \lambda}+f_{\omega\lambda}^*(\rho,\tau,\alpha)a_{\omega \lambda}^\dagger\right),
\ee
where 
\be
f_{\omega\lambda}(\rho,\tau,\alpha)=\psi_{\omega\lambda}(\rho) e^{-i\omega \tau}Y_\lambda(\alpha)
\ee
is a solution of the massive Klein-Gordon equation.  $Y_\lambda$ are the eigenfunctions of the Laplacian on $H_{d-1}$ ($\lambda$ is partly continuous and partly discrete), and $\psi_{\omega\lambda}$ behaves at large $\rho$ like\footnote{For explicit forms of $\psi_{\omega\lambda}$ and $N_{\omega\lambda}$ see appendix $A$ of \cite{Almheiri:2014lwa}.}
\be
\psi_{\omega\lambda}\to N_{\omega\lambda}\rho^{-\Delta}.  
\ee
 As in global coordinates, we can then use the extrapolate dictionary to read off that
\be
\mO_{\omega\lambda}=N_{\omega\lambda}a_{\omega\lambda}.  
\ee
Repackaging back in position space, we have finally
\be\label{rindlerrep}
\phi(x)|_{x\in W}=\int_{\partial W} dX \hat{K}(x,X)\mO(X),
\ee
where $W$ denotes the AdS-Rindler wedge and $\partial W$ denotes its intersection with the AdS boundary.  $\partial W$ is equivalent to the boundary domain of dependence (or causal diamond) of a hemisphere of the boundary timeslice at $t=0$ in global coordinates. The Rindler smearing function $\hat{K}$ is given by
\be
\hat{K}(x;\tau,\alpha)=\int_{-\infty}^\infty \frac{d\omega}{2\pi} \sum_\lambda \frac{1}{N_{\omega\lambda}}f_{\omega\lambda}(x)e^{i\omega \tau}Y_\lambda^*(\alpha).
\ee

\bfig
\includegraphics[height=5cm]{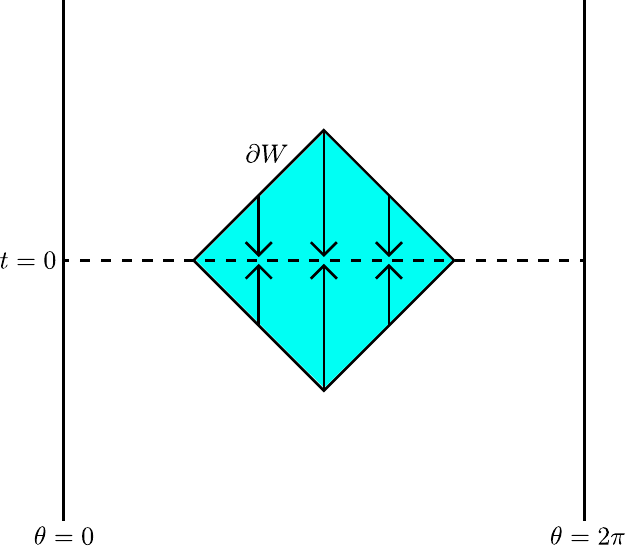}
\caption{Operator evolution to $t=0$ in the boundary of the AdS-Rindler wedge for $d=2$.}\label{rindlercollapsefig}
\efig
The key point here is that as long as a bulk scalar field lies in the AdS-Rindler wedge $W$, it has a CFT representation \eqref{rindlerrep} using only Heisenberg operators in $\partial W$.  By doing boundary evolution we can turn these into (very nonlocal) Schrodinger operators supported in the hemisphere at $t=0$, as shown for $d=2$ in figure \ref{rindlercollapsefig}.  

\bfig
\includegraphics[height=5.5cm]{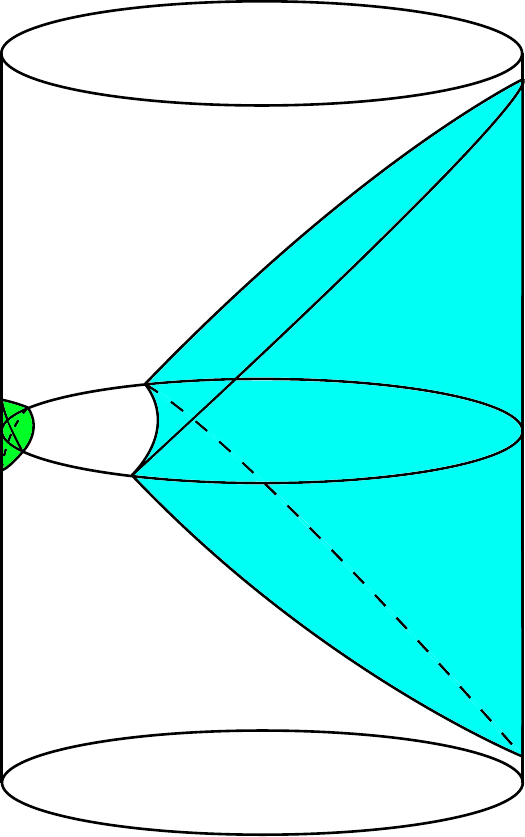}
\caption{AdS-Rindler wedges for other ball-shaped boundary regions, which for $AdS_3$ are just intervals.  In this case the ``rim'' of each wedge is a bulk geodesic connecting the endpoints of the boundary interval.}\label{multrindlerfig}
\efig
This construction can immediately be generalized from boundary hemispheres to a more general set of subregions: by acting with the conformal symmetry we can map a hemisphere of the $\mathbb{S}^{d-1}$ at $t=0$ to any other ``ball-shaped'' subregion of $\mathbb{S}^{d-1}$: we show some examples in figure \ref{multrindlerfig}.  There is an elegant covariant description of these generalized AdS-Rindler wedges, which can be stated for an arbitrary boundary spatial subregion as follows \cite{Hubeny:2012wa}:
\begin{mydef}\label{causaldef}
Let $R$ be a boundary spatial subregion of some asymptotically-AdS geometry.  We can define its boundary domain of dependence, $D[R]$, as the set of boundary points with the property that any inextendible boundary causal curve passing through one of these points must also intersect $R$.  The \textbf{causal wedge of R}, $C[R]$, is then the intersection of the bulk future and the bulk past of $D[R]$:
\be
C[R]\equiv J^+[D[R]]\cap J^-[D[R]].  
\ee
\end{mydef}
Given our program so far of bulk reconstruction by solving partial differential equations, the causal wedge is the natural guess for an upper bound on what the bulk subregion dual to the boundary subregion $R$ can be: a point which is not in the future of $D[R]$ cannot be classically affected by turning on a source in $D[R]$, and a point which is not in the past of $D[R]$ cannot be classically affected by turning on such a source and then evolving backwards in time \cite{Bousso:2012sj}.  It is more subtle to argue that fields at points which \textit{are} in $C[R]$ are indeed determined by their boundary values on $D[R]$, one argument is that a bulk observer very near $D[R]$ is able to ``send a mission'' out to any point in $C[R]$, and then receive back a report on what was going on there, which then can be immediately read out from the extrapolate dictionary \cite{Bousso:2012sj}.  Based on these arguments, and also on the explicit success of the AdS-Rindler reconstruction, we can simply conjecture that for any boundary spatial region $R$, all bulk operators in $C[R]$ can be represented in the CFT using only operators in $R$.  This is called the \textit{causal wedge reconstruction conjecture}.  Proving it along these lines would require solution of another one of those hard PDE problems: constructing the smearing function $\hat{K}$ for a generic boundary spatial region $R$.  This is complicated by the fact that even for the AdS-Rindler reconstruction, the smearing function $\hat{K}$ is not really a function, and is instead a rather delicate distribution \cite{Hamilton:2006az,Papadodimas:2012aq,Morrison:2014jha,Almheiri:2014lwa}.  In fact we will see below however that the causal wedge reconstruction conjecture can be proven for any region $R$ using completely different methods, which in fact are strong enough to allow us to reconstruct beyond the causal wedge!  The reason the upper bound argument just given breaks down is that there are other ways besides classical causal propagation for a signal to reach a point in the bulk from the boundary.   We now turn to developing the set of ideas which will lead to a complete determination of the bulk subregion dual to a boundary subregion $R$.

\section{Quantum Error Correction and Bulk Locality}\label{qecsec}
So far we have studied bulk reconstruction by solving the bulk equations of motion radially inwards from the boundary.  It would be a pity however if this were really all there were to holography.  One way to see that there must be more going on is to consider studying quantum field theory (without gravity) in an AdS background.  The Heisenberg fields in this QFT obey partial differential equations of motion, so there is nothing which stops us from using formal solutions of these equations such as eq. \eqref{iter} to express the fields anywhere in the bulk in terms of their values at the boundary.  But should we really describe this situation as holography?  Isn't gravity is supposed to be essential for the functioning of holography?  The key thing which this example misses is that, although we can indeed express the fields in the bulk in terms of their boundary values, these boundary values are not local operators in a local quantum field theory living on the boundary.  It is this boundary locality which leads to the true power of holography: it tells us that something we do not really understand (quantum gravity) is equal to something that we do understand (local quantum field theory).  In this section we explore in more detail the delicate interplay between locality in the bulk and locality in the boundary.  We will see that the difficulties in reconciling these seem at first almost insurmountable, but that once we are more careful about asking for only so much bulk locality as we really need, in the end they are compatible.  Moreover we will see that the resolution of these puzzles leads to a major expansion of the tools available for studying the bulk reconstruction problem: we can study it using the well-developed machinery of quantum error correcting codes.

\subsection{Some Puzzles}
The most obvious problem that holography needs to solve is \textit{radial locality}: how can the degrees of freedom in the center of the bulk be independent of the degrees of freedom at the boundary, given that the boundary degrees of freedom are all there is \cite{Polchinski:1999yd}?  In \cite{Almheiri:2014lwa} this problem was recast in a very sharp form: given a bulk operator $\phi(x)$ in the middle of a bulk time slice and a boundary operator $\mO(X)$ at the boundary of that timeslice, do we have
\be\label{radialc}
[\phi(x),\mO(X)]=0?
\ee
Since $x$ and $X$ are spacelike separated in the bulk, by locality in the radial direction we might expect that, at least at leading order in $G$, this commutator should indeed vanish.\footnote{At higher orders in $G$ this paradox is not so obvious, since the gravitational dressing required to make $\phi(x)$ diffeomorphism invariant reaches out to the boundary and in particular does not commute with the boundary stress tensor.  This however does not change the essential nature of the paradox: the time-slice axiom argument below would then lead us to conclude that $\phi(x)$ is a local operator with support only at the point where the dressing geodesic described below equation \eqref{dressedphi}, and this still leads to a contradiction with bulk causality \cite{Almheiri:2014lwa}.}  This equation however is extremely puzzling from the point of view of the boundary quantum field theory: aren't the local operators supposed to generate all of the other operators?  In fact there is a standard principle of mathematical quantum field theory, called the \textit{time-slice axiom}, which says as much \cite{Streater:1989vi}:\footnote{The timeslice axiom does not quite hold in field theories with topological sectors, but the loophole, which allows $B$ to be a topological surface operator, is no help here since $\phi(x)$ is clearly not topological.}
\bi
\item \textbf{Time slice axiom:} Let $\Sigma$ be any Cauchy slice, and let $U$ be any open neighborhood of $\Sigma$.  Then a bounded operator $B$ which commutes with all local operators smeared against smooth test functions compactly-supported within $U$ must obey $B\propto I$, where $I$ is the identity operator.
\ei
Intuitively, this axiom says that the set of local operators on any timeslice act irreducibly on the Hilbert space of a quantum field theory, so something which commutes with all of them must be proportional to the identity by Schur's lemma. An illustration of this idea which might be more palatable to the discretely-trained is the following: consider a spin chain of $n$ qubits.  The set of local Pauli operators on the individual qubits generate an algebra which acts irreducibly on the full Hilbert space of the $n$ qubits.  In other words, any operator on this Hilbert space can be written as a sum of products of single-site Pauli operators.  This therefore tells us that any operator which commutes with all single-site Pauli operators must commute with all operators, and must therefore be proportional to the identity.  

This argument can be viewed as a more rigorous encoding of the obvious criticism of holography: how can a quantum field theory in $d$ dimensions also be secretly local in $d+1$ dimensions, that's crazy!  In particular, if the bulk theory were a quantum field theory, such as we considered in the introduction to this section, then we really would need \eqref{radialc} to hold, and we really would be in contradiction with the time slice axiom.  There can be no AdS/CFT-like duality between quantum field theories.  In the end this is probably a good thing: quantum field theories should have the decency to know what number of spacetime dimensions they live in.  But how then is holography to realize radial locality?  Clearly gravity is going to have to be an important part of the story, but don't we also have gravity in our world, and haven't we nonetheless tested the approximate validity of quantum field theory to very high precision? 

\bfig
\includegraphics[height=4cm]{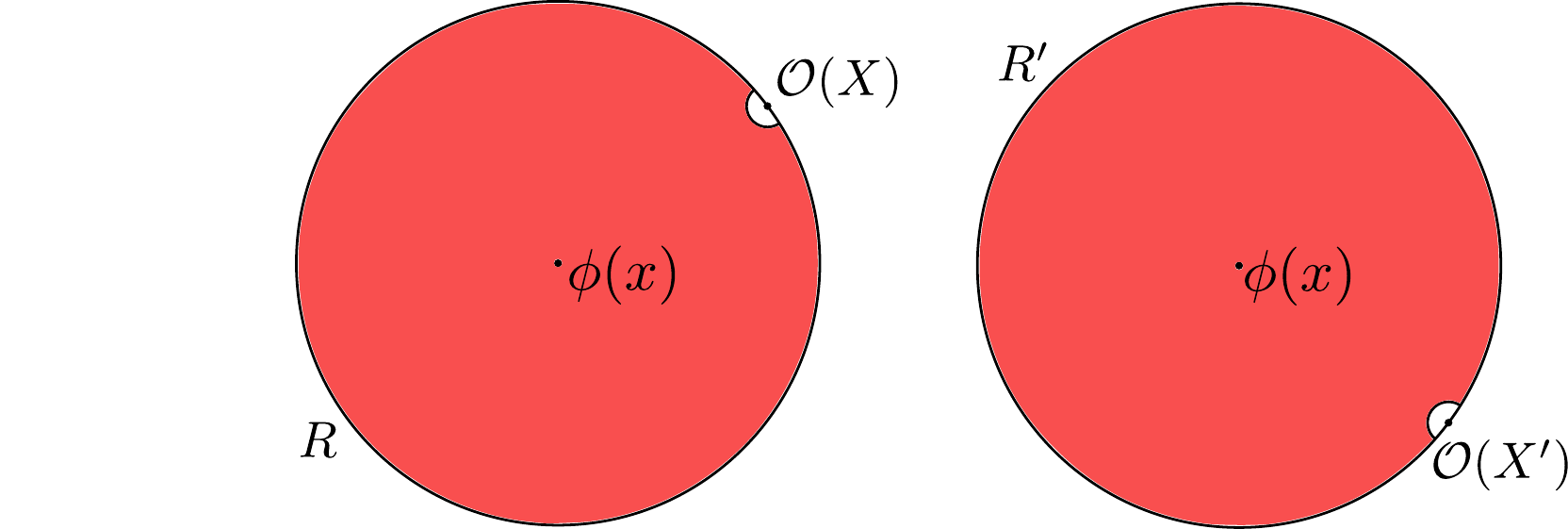}
\caption{Representing a bulk operator $\phi(x)$ on different regions to ensure a vanishing commutator with different boundary local operators.  Here we are looking at a bulk timeslice from above, and the intersections of the causal wedges $C[R]$ and $C[R']$ with this timeslice are shaded red.}\label{twoopsfig}
\efig
We can shed new light on the puzzle of radial locality using the AdS-Rindler reconstruction of the previous section.  Consider a fixed bulk operator $\phi(x)$.  By choosing an appropriate ball-shaped region on the boundary, we can arrange for an AdS-Rindler representation of $\phi(x)$ which manifestly commutes with any particular CFT local operator $\mO(X)$ which is on the same timeslice as $\phi(x)$, see figure \ref{twoopsfig}.  From the bulk side we expect these different representations to be equivalent, technically since they are just different Bogoliubov representations of the same field, but this then seems to explicitly imply that $\phi(x)$ in the CFT must commute with all $\mO(X)$, again contradicting the time slice axiom in the boundary CFT.  

\bfig
\includegraphics[height=5cm]{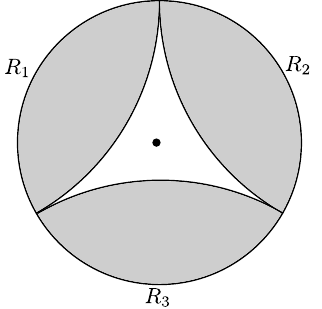}
\caption{Redundant encoding of a bulk field $\phi(x)$, which can be represented on any two of the boundary regions.}\label{triplefig}
\efig
We can make things even crazier by considering the following situation: split a boundary timeslice into three regions, $R_1$, $R_2$, and $R_3$, and consider an operator $\phi(x)$ in the center of the bulk as in figure \ref{triplefig}.  This operator does not lie in $C[R_1]$, $C[R_2]$, or $C[R_3]$, but it does lie in $C[R_1\cup R_2]$, $C[R_1\cup R_3]$, and $C[R_2\cup R_3]$.  So according to the AdS-Rindler construction, this bulk operator can be represented on \textit{any two} of the boundary regions! Where in the end is the information about this operator encoded in the boundary CFT?  This remarkable flexibility of representation appears to be in serious tension with the local structure of the degrees of freedom in quantum field theory.

The key to resolving all of these puzzles is to realize that in AdS/CFT, almost all of the states in the Hilbert space correspond to configurations in the bulk where there is a massive black hole sitting in the center of the bulk.  By contrast, the puzzles we have been discussing in this section are phrased in terms of observables such as $\phi(x)$ which we reconstructed only in the vicinity of the AdS vacuum.  We did argue that the construction should extend to states such as what we would expect near a star or something, where the backreaction could be treated by summing a subclass of Feynman diagrams,  but states where its gravitational dressing places $\phi(x)$ far beyond the event horizon of an enormous black hole are another thing altogether.  We are thus led to the idea that perhaps equations like \eqref{radialc} should be understood only as holding in a \textit{subspace} of the full CFT Hilbert space.  This idea first appeared in \cite{Papadodimas:2013jku}, who wanted to use it as part of a (controversial \cite{Harlow:2014yoa,Papadodimas:2015jra,Marolf:2015dia}) proposal to reconstruct the interiors of generic black hole microstates. Here we will instead see how it can be used as part of a new general framework for how the bulk emerges in AdS/CFT.    This framework will resolve all of the puzzles we just described, and will also settle the subregion duality question once and for all.  It is based on the idea of \textit{quantum error correction}, a subject to which we now turn.

\subsection{Three Qutrit Code}
Quantum error correction was invented to protect the memories of quantum computers from decoherence induced by interactions with the environment \cite{shor1995scheme,Gottesman:1997zz}.  The basic idea of any error correction scheme, quantum or classical, is to redundantly encode the information in a larger system so that errors are not fatal.  For example say that I wish to send you a classical bit $i$, with $i$ being zero or one.  I could just write it down on a piece of paper and send it to you, but what if the piece of paper gets lost?  Or even worse, what if some adversarial entity replaces it with a piece of paper that reads $i+1$?  The obvious way for me to protect against this is to send you the message repeatedly, for example I could send you the three-bit string $iii$.  Now if a bit is lost or flipped, as long as you do a majority vote on the bits you receive you will still be able to correctly receive the message.  This kind of protocol is called a \textit{repetition code}, it is the simplest kind of classical error-correcting code.  As with any other error-correcting code, the protection it provides is limited: if the adversary flips two of the bits we are out of luck.  

The discussion of the previous paragraph may at first seem troubling from the point of view of creating quantum error-correcting codes.  In quantum mechanics the no-cloning theorem tells us that it is impossible to make copies of a quantum state: how then are we to encode it redundantly?  Certainly the repetition code is off the table.  Quite remarkably, quantum error-correcting codes are indeed nonetheless possible.  The essential idea is to encode the information redundantly not by making copies of it, but instead by storing it in the entanglement structure of a larger number of degrees of freedom.  We will begin by illustrating this in a simple example: the three-qutrit code \cite{Cleve:1999qg}.

The three qutrit code is a quantum error-correcting code that protects the state of a single ``logical'' qutrit by storing it in a system of three ``physical''  qutrits.  Any state
\be
|\psi\ran=\sum_{i=0}^2 C_i |i\ran
\ee
of the logical qutrit is written into a three-dimensional \textit{code subspace} of the physical qutrits as
\be
|\wt{\psi}\ran=\sum_{i=0}^2 C_i |\wt{i}\ran,
\ee
with the basis states $|\wt{i}\ran$ being given by
\begin{align}\nonumber
|\wt{0}\ran&=\frac{1}{\sqrt{3}}\left(|000\ran+|111\ran+|222\ran\right)\\\nonumber
|\wt{1}\ran&=\frac{1}{\sqrt{3}}\left(|012\ran+|120\ran+|201\ran\right)\\
|\wt{2}\ran&=\frac{1}{\sqrt{3}}\left(|021\ran+|102\ran+|210\ran\right).\label{3qutrit}
\end{align}
Note that this subspace is symmetric under cyclic permutations of the physical qutrits, and that each basis state is highly entangled.  The error-correcting properties of this code arise because each basis state can be prepared via
\be\label{encode30}
|\wt{i}\ran=U_{12}\left(|i\ran_1\otimes |\chi\ran_{23}\right),
\ee
where $|\chi\ran_{23}$ indicates that the second and third physical qutrits are in the state
\be\label{chidef}
|\chi\ran\equiv \frac{1}{\sqrt{3}}\left(|00\ran+|11\ran+|22\ran\right),
\ee
and $U_{12}$ is a unitary transformation acting on physical qutrits one and two via
\begin{align}
\begin{tabular}{ l c r }
$|00\ran\to|00\ran$ & $|11\ran \to |20\ran$ & $|22\ran\to |10\ran$ \\
$|01\ran \to |11\ran$ & $|12\ran \to |01\ran$ & $|20\ran \to |21\ran$ \\
$|02\ran\to |22\ran$ & $|10\ran\to|12\ran$ & $|21\ran \to |02\ran$ \\
\end{tabular}.
\end{align}
This shows that given any logical state $|\psi\ran$, we can unitarily encode it via
\be\label{encode3}
|\wt{\psi}\ran=U_{12}\left(|\psi\ran_1\otimes |\chi\ran_{23}\right)
\ee
without any contradiction with the no-cloning theorem.  Moreover, by the cyclic symmetry of the code subspace there are also encoding unitaries $U_{31}$, $U_{23}$, with support only on the first and third or second and third physical qutrits respectively, such that 
\begin{align}\nonumber
|\wt{\psi}\ran&=U_{23}\left(|\psi\ran_2\otimes |\chi\ran_{13}\right)\\
|\wt{\psi}\ran&=U_{31}\left(|\psi\ran_3\otimes |\chi\ran_{12}\right).\label{encode32}
\end{align} 

Now we come to the main point.  Say that I want to send you a qutrit state $|\psi\ran$.  If I just send you, say,  a spin-one particle in that state, it very well might be lost on the way.  But if instead I send you three spin-one particles, which I have prepared in advance in the state $|\wt{\psi}\ran$, say by acting with $U_{12}$ as in \eqref{encode3}, the situation improves.  For example say that you only receive two out of the three particles, say the first two.  You can nonetheless take these two qutrits down to your handy quantum computer in the basement and act on them with $U_{12}^\dagger$, upon which you then have
\be
U_{12}^\dagger|\wt{\psi}\ran=|\psi\ran_1\otimes |\chi\ran_{23}.
\ee
There's the state right there in your first qutrit! And remarkably, you could have done the same thing using $U_{23}$ or $U_{31}$ had you received only the second two, or only the first and the third qutrits: the quantum state is thus protected against the loss of any one of the qutrits \cite{Cleve:1999qg}!\footnote{In this protocol it is important that you know which qutrit was lost: in quantum information parlance this means you are correcting the \textit{erasure channel}.  For subregion duality in AdS/CFT we always know which boundary region we have access to, so this is the natural type of error to discuss.}  The no-cloning theorem says only that if we can recover the state from two of physical qutrits, we had better not be able to learn anything about it from the third by itself, and indeed this is always true.  For example equation \eqref{encode3} says that for any state in the code subspace the reduced state on physical qutrit three is always maximally mixed (ie proportional to the identity), and equations \eqref{encode32} say the same for physical qutrits one and two.  

It is worth emphasizing that this redundancy of the three qutrit code relies crucially on the entanglement in the state $|\chi\ran$: say that we had an encoding map of same form as \eqref{encode3}, but with $|\chi\ran_{23}$ a product state between the second and third physical qutrits.  This would mean that in the encoded state, the third qutrit was simply sitting in the same pure state for all states in the code subspace: there would be no way for the second and third qutrits together to have any information about the encoded message which the second qutrit did not have by itself.  

This correctability should be somewhat reminiscent of the bulk operator representable on any two of three boundary subregions that we illustrated in figure \ref{triplefig}, and we can make it even more similar by introducing the notion of \textit{logical operators}.  A logical operator is just any linear map on the logical qutrit:
\be\label{Odef}
O|i\ran=\sum_j(O)_{ji}|j\ran.
\ee
Any logical operator can be lifted to an encoded logical operator which acts on the code subspace via
\begin{align}\nonumber
\wt{O}|\wt{i}\ran&=\sum_j(O)_{ji}|\wt{j}\ran\\
\wt{O}^\dagger|\wt{i}\ran&=\sum_j(O)_{ij}^*|\wt{j}\ran,
\end{align}
with the action of $\wt{O}$ on the orthogonal complement of the code subspace left arbitrary, but for an arbitrary code subspace we would expect that such an encoded logical operator would usually need to have support on all three of the physical qutrits.  The situation is better however for the particular code subspace given by \eqref{3qutrit}.   Indeed the operator
\be\label{O12}
O_{12}\equiv U_{12}O_1 U_{12}^\dagger,
\ee
where $O_1$ is an operator which acts on the first physical qutrit with matrix elements $(O)_{ij}$,\footnote{Note that in $O_{12}$ and $U_{12}$, the ``12'' is a label telling us which physical qutrits the operator acts on, it is not the matrix indices which appear in \eqref{Odef}. The latter will never appear again, but anyways to avoid a complete notational clash I wrote the matrix elements as $(O)_{ij}$.  This situation is not improved by the great pedagogical tragedy of the three qutrit code, which is that each qutrit has three states and there are also three such qutrits, but I hope the reader can manage.} acts on any state $|\wt{\psi}\ran$ in the code subspace as
\begin{align}\nonumber
O_{12}|\wt{\psi}\ran&=\wt{O}|\wt{\psi}\ran\\
O_{12}^\dagger|\wt{\psi}\ran&=\wt{O}^\dagger |\wt{\psi}\ran.
\end{align}
Thus any logical operator can be represented on only the first two of the physical qutrits, and in fact using $U_{23}$ and $U_{31}$ it can also be represented on the second and third or first and third physical qutrits.  

\bfig
\includegraphics[height=5cm]{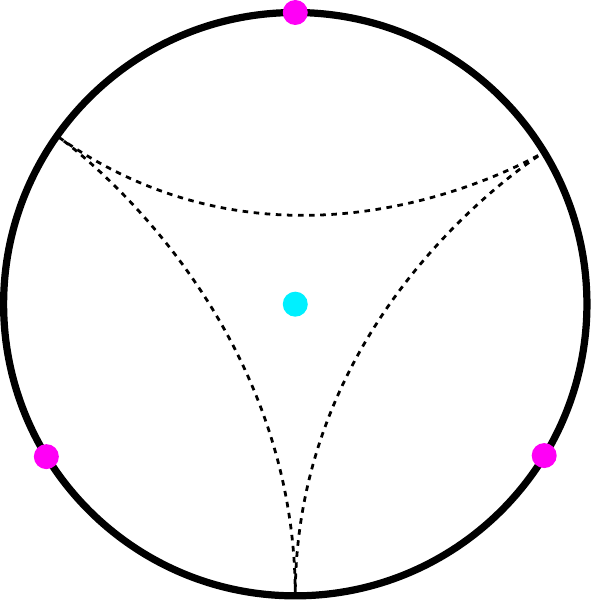}
\caption{The three qutrit model of holography: one ``bulk effective field theory'' qutrit is encoded into three ``boundary CFT'' qutrits.}\label{qutritsfig1}
\efig
This is getting quite close to the situation of figure \ref{triplefig} in the previous section, let's now make the analogy explicit:
\bi
\item The three physical qutrits are the local degrees of freedom of a boundary ``CFT'' with three lattice sites.  Their 27-dimensional Hilbert space is the full Hilbert space of this ``CFT''.  
\item The one logical qutrit is the local degree of freedom of a bulk ``effective field theory'' with one lattice site, with that lattice site sitting in the center of the bulk.  Logical operators are interpreted as bulk fields at that point. 
\ei   
We illustrate this in figure \ref{qutritsfig1}.  We can think of the bulk point is lying in the causal wedge of any two of the boundary qutrits, and our construction of $O_{12}$, $O_{23}$, and $O_{31}$ can be interpreted as the AdS-Rindler reconstruction.  Moreover following the logic of figure \ref{twoopsfig}, we can see that radial locality holds in the following sense: let $\wt{O}$ be any encoded logical operator, and $X_3$ be any operator acting only on the third physical qutrit.  For any two states $|\wt{\psi}\ran$, $|\wt{\phi}\ran$ in the subspace, we have
\begin{align}
\lan\wt{\phi}|[\wt{O},X_3]|\wt{\psi}\ran=\lan\wt{\phi}|[O_{12},X_3]|\wt{\psi}\ran=0.
\end{align}
Moreover the same is true for any operators $X_1$ on the first physical qutrit or $X_2$ on the second physical qutrit, since we can instead replace $\wt{O}$ by $O_{23}$ or $O_{31}$.  Thus between states in the code subspace, we can indeed have nontrivial bulk field operators that commute with all local operators in the ``boundary CFT''.  There is no tension with the timeslice axiom, which is certainly true for three qutrits, because have asked for the analogue of equation \eqref{radialc} to hold \textit{only in the code subspace}.  

\bfig
\includegraphics[height=5cm]{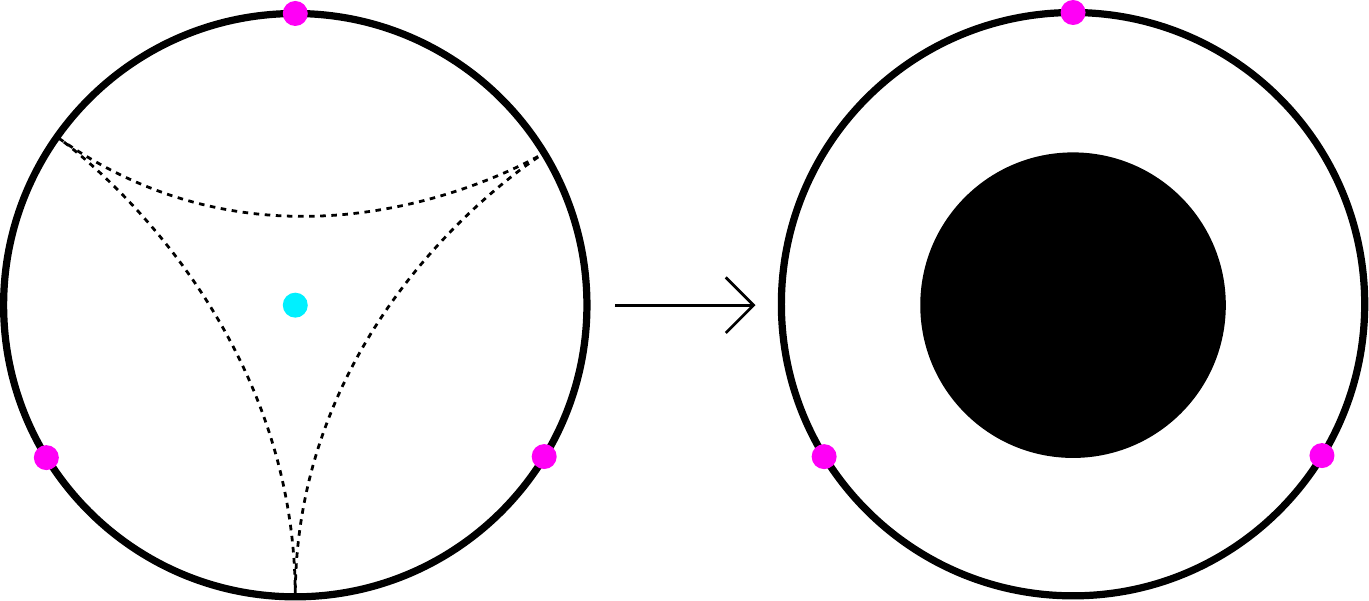}
\caption{Black holes in the three qutrit code: as we move out of the code subspace we transition from bulk effective field theory near the vacuum to a big black hole which has swallowed our only bulk point.  This black hole has $\log_3 24$ qutrits worth of microstates.}\label{qutrits2fig}
\efig
This immediately begs the question however of what happens in the orthogonal complement of the code subspace: what about rest of the Hilbert space?  AdS/CFT is suppose to be an isomorphism between the bulk and the boundary, so these states had better have bulk interpretations as well.  Indeed they do, as shown in figure \ref{qutrits2fig} they are accounted for by the microstates of a big black hole!  This claim may seem a bit arbitrary, after all the Hilbert space only has 27 dimensions so who is to say what is a black hole and what isn't, but we will see in the following two sections that this interpretation generalizes in a natural way to situations where it is clearly correct.  

The three qutrit model of holography really does quite well for such a trivial model: bulk effective field theory, radial locality, AdS-Rindler Reconstruction, and black hole entropy all have plausible avatars.  Moreover we will see soon that the Ryu-Takayanagi formula also holds in this model.  At least for the author it has been very useful to be able to probe these ideas in such a simple context.  Nonetheless it would obviously be nice to have models which contain a bit more of the structure of full AdS/CFT.  In the following two subsections we first understand a bit more about what structures we should hope for, and then indeed construct a model which possesses them.  

\subsection{General Properties of Holographic Codes}
There is an elegant general theory of quantum error correcting codes, which we unfortunately do not have time to explore in detail \cite{Gottesman:1997zz,nielsen2002quantum}, but there are a few nice results which are worth discussing.  In this subsection we will first describe these results, and then apply them to holography to understand some general features of the holographic code.  

The first result we'll discuss is a theorem \cite{Schumacher:1996dy,Grassl:1996eh} which says that the structure \eqref{encode3} of the three qutrit code subspace is actually a general feature of quantum codes in which all information in the code subspace is accessible from a subset of the physical degrees of freedom:
\begin{thm}\label{standardthm}
Let $\mathcal{H}$ be a finite-dimensional Hilbert space which tensor factorizes into $\mathcal{H}_R\otimes \mathcal{H}_{\ol{R}}$, and let $\mathcal{H}_{code}$ be a subspace of $\mathcal{H}$.  Then the following are equivalent:
\bi
\item For any operator $\wt{O}$ on $\Hc$, there exists an operator $O_R$ on $\Hh_R$ such that for all $|\wt{\psi}\ran\in \Hc$ we have
\begin{align}\nonumber
O_R|\wt{\psi}\ran&=\wt{O}|\wt{\psi}\ran\\
O_R^\dagger|\wt{\psi}\ran&=\wt{O}^\dagger|\wt{\psi}\ran.
\end{align}
\item For any operator $X_{\ol{R}}$ on $\Hh_{\ol{R}}$ we have 
\be
PX_{\ol{R}}P=\lambda P,
\ee
where $P$ is the projection onto $\Hc$ and $\lambda\in \mathbb{C}$.
\item If we introduce a reference system $S$ whose dimensionality equals that of $\Hc$, then the state
\be\label{phidef}
|\phi\ran\equiv \frac{1}{\sqrt{|S|}}\sum_i|i\ran_S |\wt{i}\ran_{R\ol{R}},
\ee
where $|\wt{i}\ran_{R\ol{R}}$ and $|i\ran$ are bases for $\Hc$ and $\Hh_S$ respectively, has the property that
\be
\rho_{S\ol{R}}(\phi)=\rho_{S}(\phi)\otimes \rho_{\ol{R}}(\phi).
\ee
\item $|S|\leq |R|$, and if we decompose $\Hh_R=(\Hh_{R_1}\otimes\Hh_{R_2})\oplus \Hh_{R_3}$, with $|R_1|=|S|$ and $|R_3|<|S|$ (this is always possible by long division), then there exists a unitary operator $U_R$ on $\Hh_R$ and a state $|\chi\ran_{R_2\ol{R}}$ on $R_2\ol{R}$ such that
\be\label{standardencode}
|\wt{i}\ran_{R\ol{R}}=U_R\left(|i\ran_{R_1}|\chi\ran_{R_2\ol{R}}\right).
\ee
\ei
\end{thm}
Each of these conditions quite plausibly is necessary for an erasure of $\ol{R}$ to be correctable: (1) says all logical operators can be represented on $R$, (2) says that no measurement on $\ol{R}$ can tell us about the encoded message, (3) says that there is no correlation between $\ol{R}$ and the encoded information, and (4) tells us that we can recover the state as in \eqref{encode3}.  The encoding structure promised by condition (4) is nicely represented as a circuit diagram, shown in figure \ref{circuit1fig}.
\bfig
\includegraphics[height=4cm]{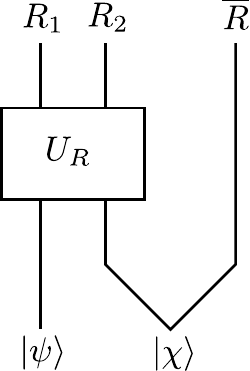}
\caption{Encoding a standard erasure code, as in equation \eqref{standardencode}.}\label{circuit1fig}
\efig The proof of theorem \ref{standardthm} is instructive, and not particularly difficult, but since it is presented in section 3.1 of \cite{Harlow:2016vwg} I'll omit it here.  

We can now use theorem \ref{standardthm} to derive the \textit{quantum singleton bound} \cite{Knill:1996ny}, which in certain symmetric situations quantifies how large the code subspace can be while still preserving correctability.  Say that we have a Hilbert space which tensor factorizes into a set of $n$ physical qudits (a qudit is a $d$-state qubit), and that we want to encode $k$ logical qudits into a $d^k$-dimensional code subspace.  In general we cannot say much about how big $k$ can be relative to $n$, but we can do much better if we assume that whether or not a collection of physical qudits is able to access the logical information is determined entirely by the number of qudits in that collection.  Indeed say that $m$ is the smallest number of qudits from which we may still access the information in the sense of theorem \ref{standardthm}.  It is clear that $m$ must be greater than $n/2$, since otherwise we could clone quantum information by recovering it on each half of the system.  We may therefore decompose the system into three parts $A$, $B$, and $C$, where $A$ and $B$ each consist of $n-m$ qudits and $C$ is non-empty.  The quantum singleton bound is derived by noting that since the erasure of either $A$ or $B$ (but not both) is correctable, by theorem \ref{standardthm} the state $|\phi\ran_{ABCS}$ obtained by entangling the code subspace with a reference system $S$, as in \eqref{phidef}, must obey\footnote{This argument uses various properties of von Nuemann entropy and pure states, readers for whom these are unfamiliar may wish to refer to appendix C of \cite{Harlow:2014yka}.}
\begin{align}\nonumber
\rho_{AS}(\phi)&=\rho_A(\phi)\otimes \rho_S(\phi)\\
\rho_{BS}(\phi)&=\rho_B(\phi)\otimes \rho_S(\phi),
\end{align}
and therefore the von Neumann entropies of these subsystems must obey
\begin{align}\nonumber
S_{AS}&=S_A+k\log d\\
S_{BS}&=S_B+k\log d.
\end{align}
Moreover by subadditivity of von Neumann entropy we must have
\begin{align}\nonumber
S_{AC}&\leq S_A+S_C\\
S_{BC}&\leq S_B+S_C.
\end{align}
We therefore have
\begin{align}\nonumber
k\log d&=S_{AS}-S_A=S_{BC}-S_A\leq S_C+S_B-S_A\\
k\log d&=S_{BS}-S_B=S_{AC}-S_B\leq S_C+S_A-S_B.
\end{align}
These inequalities cannot both be satisfied unless $k\log d\leq S_C\leq (2m-n)\log d$, so we thus arrive at the quantum singleton bound
\be\label{singleton}
m\geq \frac{n+k}{2}.
\ee
This result is quite plausible: to recover a bigger message, we need access to more qudits.  Roughly speaking, if $k$ is small compared to $n$ then we need at least half of the qudits, while if they are comparable we need more. As a simple check, in the three qutrit code we have $n=3$, $k=1$, and $m=2$, and indeed we have $1+3=2\times 2$.  In fact not only is \eqref{singleton} necessary for the correctability of the code (given our symmetry assumption), it is also typically \textit{sufficient}: if we choose the code subspace randomly, then as we satisfy \eqref{singleton} better and better, the code becomes correctable up to errors that are exponentially small in $2m-n-k$ \cite{Almheiri:2014lwa}.    

We'll now apply these results to holography.  We first need to pick a code subspace.  There is no unique choice, different choices are possible which highlight different physics.  This is somewhat reminiscent of the choice of renormalization scale in effective field theory, and indeed we will see below that these two choices are closely related.  For now a simple choice that leads us already to interesting physics is to consider the linear span of the set of states obtained from the vacuum by acting with of order $k$ bulk operators which lie within a region in the center of the bulk whose size is small but $O(N^0)$ in AdS units.  These bulk operators may be represented in the CFT using global reconstruction, as in equation \eqref{phi3eq}.  To avoid UV singularities we can smear these bulk operators against test functions of compact support within this central region, which do not vary much over scales that are parametrically small compared to the region.  Since this region is in the center of $AdS$, symmetry says that whether or not any particular ball-shaped region on the boundary can access the logical information is determined only by the radius of that ball-shaped region.  Since the complement of a ball-shaped region is also a ball-shaped region, it turns out that we have enough symmetry that a version of the argument which established the quantum singleton bound also applies here.  Namely we can consider two disjoint ball-shaped regions $A$ and $B$, each of which has the maximal radius such that the logical information in the center can be accessed from its complement.  The same argument as for the quantum singleton bound\footnote{We can't apply the singleton bound directly, since here we have only assumed the symmetry for ball-shaped regions rather than arbitrary collections of qudits, but the derivation generalizes immediately.} now tells us that we must have
\be\label{cftsing}
\log \mathrm{dim}(\Hc)\leq S_C.
\ee
The dimensionality of the code subspace is of order $e^{\alpha k}$, with $\alpha$ some $O(N^0)$ constant which depends in detail on how we smear the operators and where we allow them to be placed.  $S_C$ of course will be infinite, since it is the entropy of a subregion in a quantum field theory, but in fact most of that entropy is associated to short-distance degrees of freedom which are fixed for all states in the code subspace, and which are thus irrelevant for the problem at hand.  More precisely, the smearing function $K$ does not vary much as a function of the boundary point when the bulk point is in the center, so we may therefore integrate out most of the degrees of freedom in the CFT, down to a lattice with some large but $O(N^0)$ number of points on the spatial sphere.  At each of these points, the discussion above equation \eqref{Scft} suggests that we should have of order $N^{\sharp}$ degrees of freedom, with $\sharp$ defined in equation \eqref{Geq}.  For any region $R$ in this regularized Hilbert space we therefore have
\be
S_R\leq  \beta N^\sharp V_R,
\ee
where $\beta$ is some other $O(N^0)$ constant and $V_R$ denotes the number of cells in $R$.  In particular if we denote by $V_{min}$ the cell volume of the smallest ball from which we can access the information, and by $V_{sphere}$ the total number of cells, then we have $V_C=V_{sphere}-2V_{min}$, so \eqref{cftsing} tells us that
\be\label{volbound}
V_{min}\geq \frac{V_{sphere}+\alpha \beta^{-1}N^{-\sharp}k}{2}. 
\ee

\bfig
\includegraphics[height=4cm]{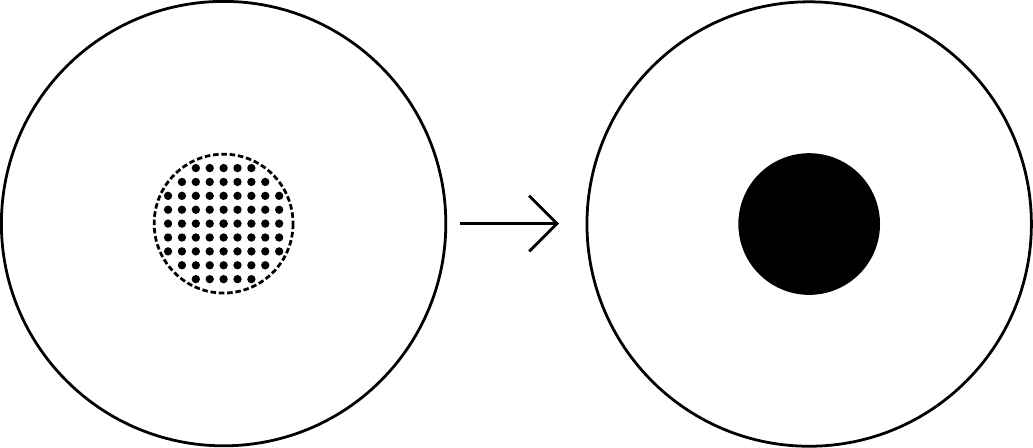}
\caption{The essential mechanism of holography: including too many operators in the code subspace leads to black hole formation and a breakdown of correctability.}\label{holpicfig}
\efig
Let us now first consider the case where $k\sim N^0$.  This code subspace describes a few perturbative excitations about the vacuum, localized near the center of the bulk.  The second term in the numerator of \eqref{volbound} may then be ignored, so \eqref{volbound} tells us that we need a ball with at least half of the boundary lattice sites to access these degrees of freedom.  And that is indeed what the AdS-Rindler reconstruction tells us: to represent bulk operators in the center of the bulk, we need at least half of the boundary.    We may now start increasing $k$.  For a while nothing interesting happens, since the other term in \eqref{volbound} is still small.  But eventually we reach $k\sim N^{\sharp}$, and the correctability of the code will begin to fail.  But from equation \eqref{Sbulk} this is precisely when we \textit{expect} it to fail: it is when most of the states we create this way correspond to a large black hole sitting in the center of the bulk!\footnote{Note that this argument was not able to determine the $O(1)$ coefficient in the entropy of the black hole: this is buried in the constants $\alpha$ and $\beta$ which we did not compute.}   See figure \ref{holpicfig}.  We can view this as the essential mechanism of holography: we try as hard as we can to force the CFT to admit that it is really just a $d$ dimensional quantum field theory, but right when it will no longer be able to continue realizing locality in the emergent radial direction, black hole formation comes in to save the day. 

If now we act on this code subspace with a conformal transformation to move the logical information around, we see another interesting consequence of the quantum error correction interpretation of AdS/CFT: there is a simple relationship between location in the radial direction and degree of protection from erasures.  We illustrate this basic point in figure \ref{erasurefig}.  Namely if we lose access to a ball-shaped region in the boundary, we also lose access to all bulk operators in its causal wedge.  In particular the larger the erasure, the deeper into the bulk the corruption goes.  This is a precise reformulation of the old somewhat fuzzy notion of ``scale-radius duality'' in AdS/CFT \cite{Susskind:1998dq,deBoer:1999tgo,Heemskerk:2010hk}.
\bfig
\includegraphics[height=5cm]{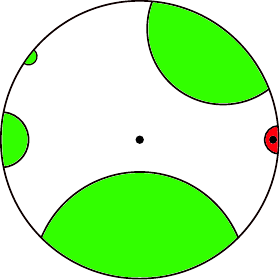}
\caption{Operators near the center of the bulk are protected from rather large erasures, (I've shaded a few such erasures green), while operators near the boundary are easily corrupted by small erasures (eg the one I've shaded red).}\label{erasurefig}
\efig

We will momentarily illustrate these somewhat formal arguments in a concrete example, but before doing so I need to point out that the situation described by theorem \ref{standardthm} is not general enough to accommodate some of the more interesting code subspaces we might study.  For example if we wish to exclude black holes, one very natural choice of code subspace is the set of CFT states whose energy is less than some large but $O(N^0)$ energy. For example in the $\mathcal{N}=4$ super Yang Mills theory with gauge group $SU(N)$ and 't Hooft coupling $\lambda$ large but $O(N^0)$, we might consider the states obtained through the state-operator correspondence from all primary and descendant operators with conformal dimension at most of order $\lambda^{1/4-\epsilon}$, for some $\epsilon>0$.  The tensor network code subspace we meet in the next subsection will basically be analogous to this subspace.  But we are not able to describe the emergence of the bulk in such a code subspace using an encoding of the form \eqref{standardencode}.  The basic problem is that on this code subspace the logical operators include low energy bulk operators anywhere in the space (except for parametrically close to the boundary where they create too much energy), and \eqref{standardencode} enables us to represent \textit{all} logical operators on the boundary region $R$.  This is not what we would expect from subregion duality, which would say that we should instead expect to realize only those bulk operators which are in the causal wedge of $R$ (or really the ``entanglement wedge'', see below).  So far I avoided this by considering only a code subspace and a region where all the relevant operators were indeed in the causal wedge.  But this is clearly too restrictive, for example in figure \ref{erasurefig} it is natural to treat both of the bulk local operators as logical operators on the same code subspace, as indeed they would be in the ``low-energy'' code subspace I just described.  But if we demand the ability to represent both of these operators on a boundary region $R$, this will be quite constraining on $R$, since we would need to require that the causal wedge $C[R]$ contain both operators.  Indeed one of the whole points of quantum error correction is that we should pick our code subspace before we know what the errors are.  Fixing this leads us to the notion of \textit{subsystem quantum error correction}, where given any boundary region $R$ we ask only for representations of those logical operators which live in an appropriate subregion of the bulk.  The basic idea is that if the code subspace factorizes into 
\be
\Hc=\mathcal{H}_r\otimes \mathcal{H}_{\ol{r}},
\ee
with a complete basis $|i\ran_r |j\ran_{\ol{r}}$, then we can have a generalization of equation \eqref{standardencode} where the encoding map is
\be\label{compenc}
|\wt{ij}\ran=U_R U_{\ol{R}}\left(|i\ran_{R_1} |j\ran_{\ol{R}_1}|\chi\ran_{R_2\ol{R}_2}\right).
\ee
We can think of $\mathcal{H}_r$ as the bulk degrees of freedom in $C[R]$ and $\mathcal{H}_{\ol{r}}$ as the bulk degrees of freedom in $C[\ol{R}]$, although we will see soon that we really mean the degrees of freedom in their ``entanglement wedges'' instead of their causal wedges (these coincide for ball-shaped regions in pure $AdS$).  In \cite{Harlow:2016vwg} I called the situation described by \eqref{compenc} ``subsystem quantum error correction with complementary recovery'', and I proved a theorem analogous to theorem \ref{standardthm} saying that the encoding must take this form in any situation where all logical operators on $\mathcal{H}_r$ can be represented on $R$ and all logical operators on $\mathcal{H}_{\ol{r}}$ can be represented on $\ol{R}$.  We will see this in action in the following subsection.

\subsection{A Tensor Network Model of Holography}
I'll now describe a generalization of the three-qutrit code which illustrates the features of the holographic correspondence discussed in the previous subsection \cite{Pastawski:2015qua}. This code has received quite a bit of attention, so some historical background may be of interest.  After writing \cite{Almheiri:2014lwa} with Xi Dong and Ahmed Almheiri, I went to Caltech in the fall of 2014 to give a seminar.  I presented more or less the material which has been described in this section so far, and I closed by saying that I thought there should be some MERA-like tensor network which could explicitly realize what we had described (this comment was motivated by the landmark papers \cite{Swingle:2009bg,Qi:2013caa}).  Fortunately for me, Fernando Pastawski and Beni Yoshida were in the audience.  I went back to the Athenaeum for the evening, and the next morning they came to find me having more or less constructed the final version of the code which I describe in this section.  John Preskill and I helped fill in the rest, and together we wrote \cite{Pastawski:2015qua}.  We immediately found ourselves in a quandary however: although in high-energy physics author lists are almost always alphabetical, in this case it was clear to all of us that we should acknowledge the key role played by Fernando and Beni.  The problem however was that doing this would destroy one of the best author acronyms any of us had ever encountered.  After agonizing about this for several weeks, we decided that in the end we simply had to feature Fernando and Beni \cite{Pastawski:2015qua}.  I am glad we did.  Nonetheless it was a pity to lose the acronym, so we started to use it anyways to colloquially refer to the construction: thus the HaPPY code was born.

\bfig
\includegraphics[height=4cm]{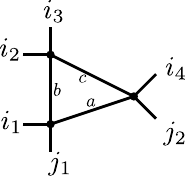}
\caption{A simple tensor network, which represents the tensor $T_{i_1i_2i_3i_4,j_1j_2}=\sum_{a,b,c}T_{i_1j_1ab}T_{i_2i_3bc}T_{i_4j_2ac}$.}\label{networkfig}
\efig
The HaPPY code is an error correcting code which encodes $k$ ``bulk'' logical qubits into $n$ ``boundary'' physical qubits, with the encoding map analogous to eq. \eqref{3qutrit} above given by an $n+k$-index tensor
\be\label{HaPPYencode}
\lan i_1\ldots i_n|\wt{j_1\ldots j_k}\ran\equiv T_{i_1\ldots i_n,j_1\ldots j_k}.
\ee
This tensor is generated by a \textit{tensor network}, which is a quite general way of constructing complicated tensors with many indices out of smaller tensors with fewer indices. The basic idea is to represent the tensor as a graph whose vertices represent the smaller tensors: the edges attached to each vertex represent indices of the tensor at that vertex.  Edges which connect two vertices represent index contraction, while edges which are attached to only one vertex represent free indices of the final tensor. I give a simple example in figure \ref{networkfig}.

The building blocks for the HaPPY code are what we called \textit{perfect tensors}: these are tensors with an even number of indices which have the property that any balanced bipartition of the indices into inputs and outputs gives a unitary transformation.    It is not obvious that such tensors exist, but they do and in fact we have already met one: the encoding map for the three qutrit code,
\be
T_{i_1 i_2 i_3, j}\equiv \sqrt{3}\lan i_1i_2i_3|\wt{j}\ran,
\ee
with the states $|\wt{j}\ran$ given by \eqref{3qutrit}.  To see that this tensor is perfect, note that using \eqref{encode30} and \eqref{chidef}, we have 
\begin{align}\nonumber
T_{i_1 i_2 i_3, j}&=\sqrt{3}\lan i_1 i_2 i_3|U_{12} |j\ran|\chi\ran\\\nonumber
&=\sum_k \lan i_1 i_2 i_3|U_{12} |j\ran|k\ran|k\ran\\
&=\lan i_1 i_2|U_{12}|j i_3\ran.
\end{align}
This shows that if we view the first pair of indices of $T$ as outputs and the second pair as inputs, $T$ is unitary.  Unitarity for the other balanced bipartitions follows from equations \eqref{encode32}. Indeed perfect tensors are always associated to quantum error correcting codes.  To build the HaPPY code, the perfect tensor for the three qutrit code wasn't good enough so we instead used the perfect tensor for another standard quantum error correcting code: the five qubit code \cite{Gottesman:1997zz}
\be
T_{i_1i_2i_3i_4i_5,j}\equiv 2\lan i_1 i_2i_3i_4i_5|\wt{j}\ran.
\ee
This code encodes one logical qubit into five physical qubits, and protects against the erasure of any two of the qubits.  Note that since $2\times3=5+1$, this is consistent with the quantum singleton bound \eqref{singleton}.  The details of this code will not matter for us, but we note that the code subspace is defined as the two-dimensional subspace of the 32-dimensional Hilbert space of the five physical qubits which has eigenvalue one under the ``stabilizer'' operators 
\begin{align}\nonumber
S_1&=X\otimes Z\otimes Z\otimes X\otimes I\\
S_2&=I\otimes X \otimes Z \otimes Z \otimes X\\\nonumber
S_3&=X\otimes I \otimes X \otimes Z \otimes Z\\\nonumber
S_4&=Z\otimes X \otimes I \otimes X \otimes Z.
\end{align}
Here I have adopted the quantum information convention that the Pauli operators are written as $X,Y, Z$.  This subspace is symmetric under cyclic permutations of the five physical qubits.  
\bfig
\includegraphics[height=2cm]{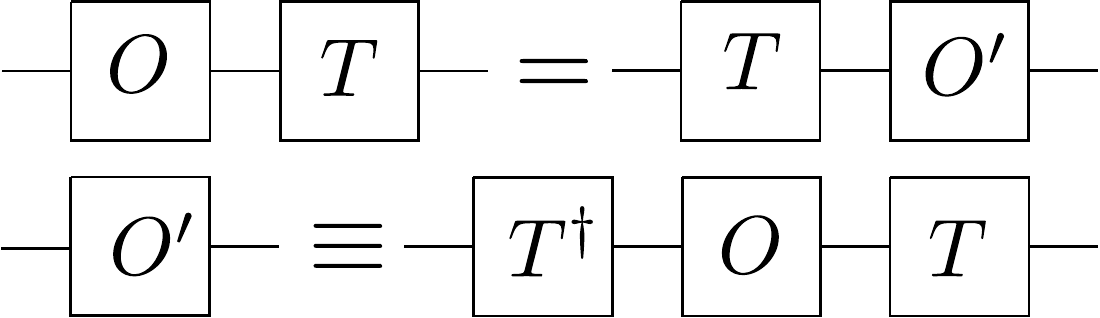}
\caption{Operator pushing.}\label{pushfig}
\efig
Perfect tensors have an important ``operator pushing'' property: an operator which acts on up to half of the indices of a perfect tensor can be replaced by an operator acting on the remaining indices.  This is illustrated in figure \ref{pushfig}.

\bfig
\includegraphics[height=8cm]{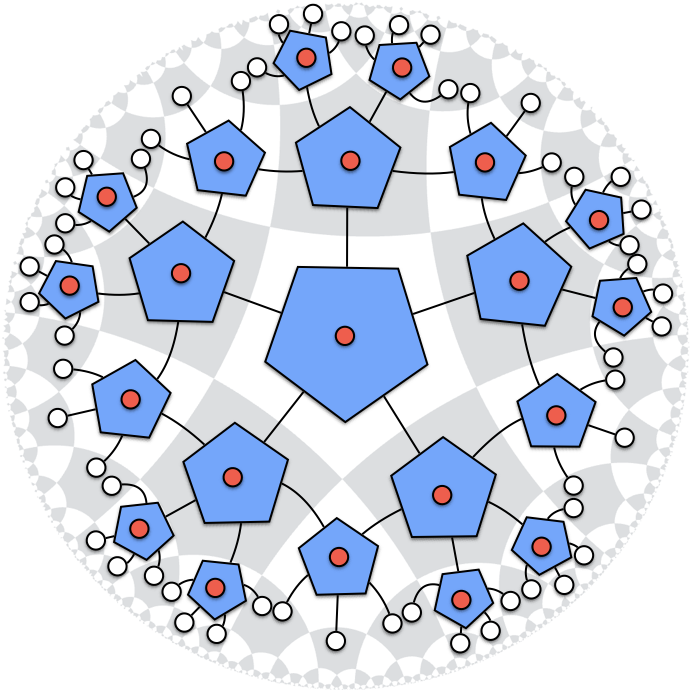}
\caption{The HaPPY tensor network.  The red dots denote logical ``bulk'' indices, while the white dots denote physical ``boundary'' indices.  A ``continuum'' limit may be taken as we extend the network all the way to the boundary.  }\label{happyfig}
\efig
The HaPPY encoding map is then constructed as follows: we begin with a standard pentagon tiling of the hyperbolic plane, and insert a six-qubit perfect tensor in the center of each pentagon.  The ``logical'' leg of each tensor is left free, and corresponds to one of the logical bulk $j$ indices in \eqref{HaPPYencode}.  The five ``physical'' legs of each of the perfect tensors are then contracted with their neighbors through the edges of the tiling.  We then regulate the tiling at some large but finite radius, and the legs through which the regulating surface passes become the physical boundary indices of \eqref{HaPPYencode}.  This is illustrated in figure \ref{happyfig}.  I won't give the detailed argument here, but this network defines a \textit{isometric embedding}, in the sense that the image through the network of an orthonormal basis for the bulk indices is an orthonormal basis of the code subspace.  The real point however is that we can use the operator pushing operation of figure \ref{pushfig} to construct a version of the AdS-Rindler reconstruction for the HaPPY code: this is illustrated in figure \ref{pentagonpushfig}.

\bfig
\includegraphics[height=8cm]{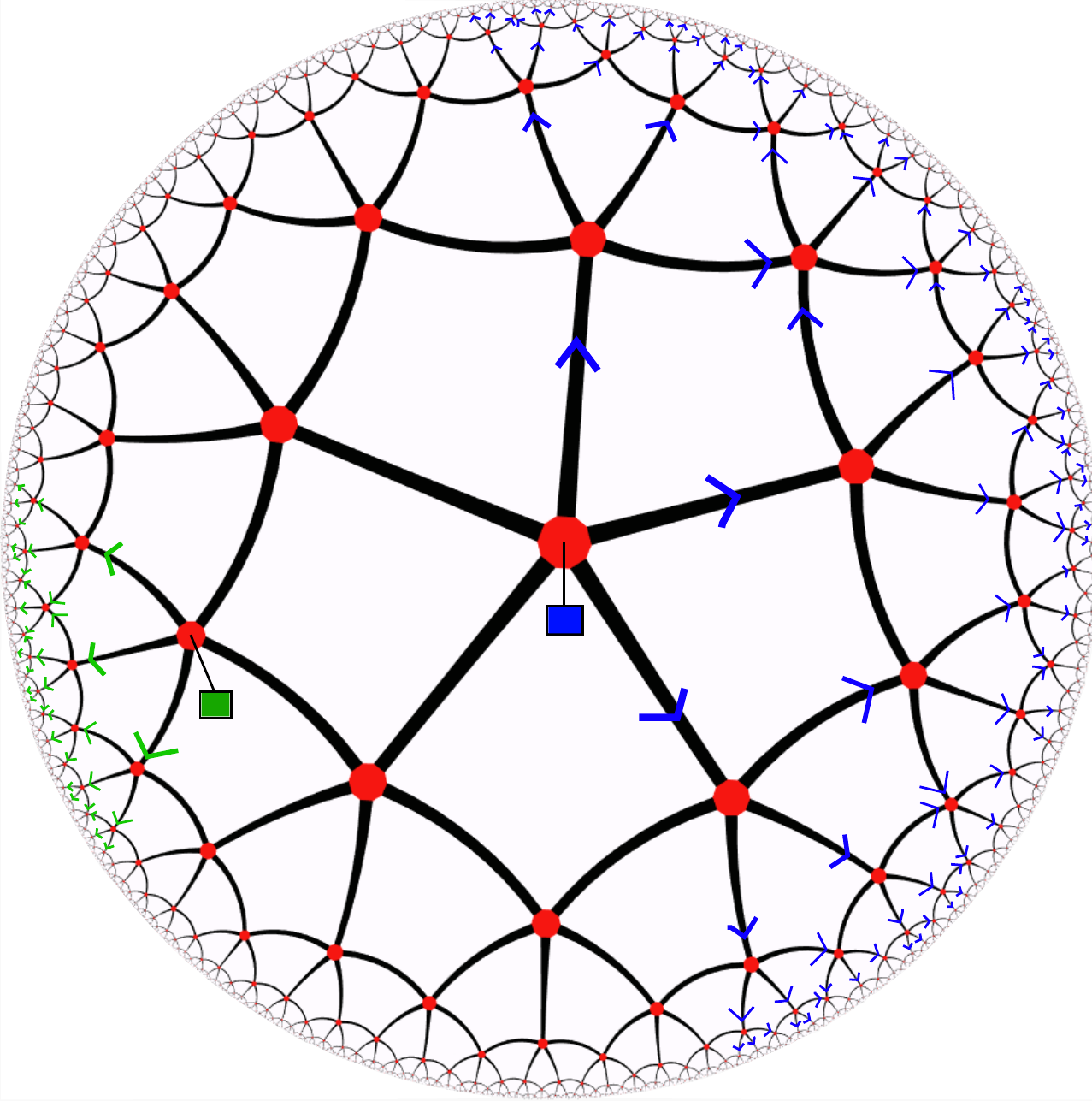}
\caption{AdS-Rindler reconstruction in the HaPPY code.  An operator in the center of the bulk can only be reconstructed using a large fraction of the boundary, while an operator closer to the boundary can be reconstructed on a smaller fraction.  We have quite a bit of choice how to push operators through the network, which is a graphical representation of the redundancy of the code.  This figure has become the logo of the Simons Foundation ``It from Qubit'' Collaboration.}\label{pentagonpushfig}
\efig  
We now have within the code subspace a volume's worth of bulk degrees of freedom, each of which commutes with all boundary local operators in matrix elements within the code subspace, and thus satisfies radial locality.  Moreover, figure \ref{pentagonpushfig} makes it obvious that the AdS-Rindler reconstruction of the HaPPY code obeys the relationship between radial location and protection from erasures outlined in the previous subsection.   We can formalize the structure of the code using the language of subsystem quantum error correction with complementary recovery introduced at the end of the last section, as shown in figure \ref{cutnetworkfig}.

\bfig
\includegraphics[height=7cm]{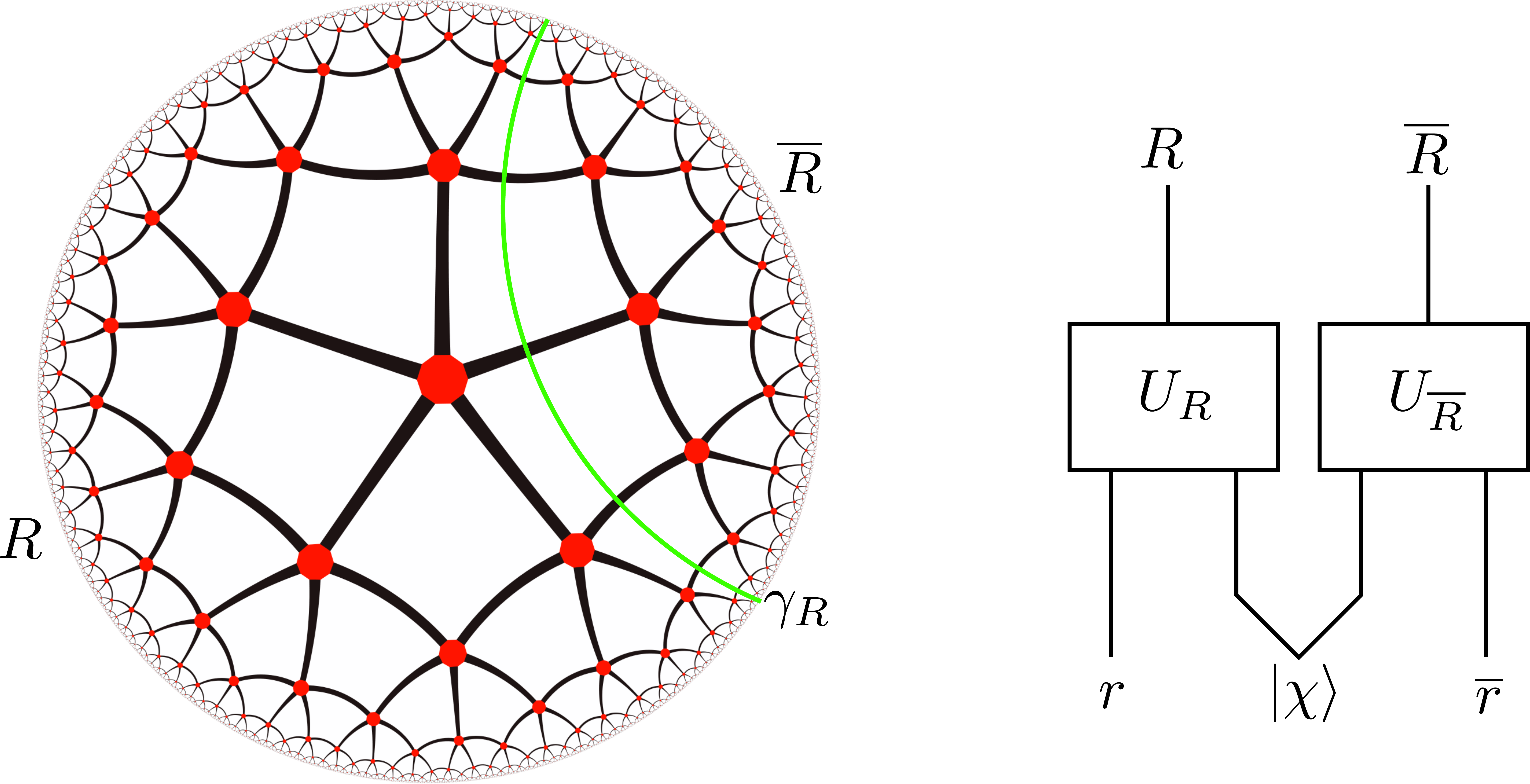}
\caption{Comparing the HaPPY network to the encoding circuit for subsystem quantum error correction with complementary recovery, as defined via \eqref{compenc}.  The green line $\gamma_R$ divides the bulk into the sites $r$ from which operators can be pushed to $R$ and the sites $\ol{r}$ from which they can be pushed to $\ol{R}$.  In the encoding map the state $|\chi\ran$ corresponds to the links which are cut by the green line, the unitary $U_R$ is the piece of the tensor network between $\gamma_R$ and $R$, and the unitary $U_{\ol{R}}$ is the piece of the tensor network between $\gamma_R$ and $\ol{R}$.}\label{cutnetworkfig}
\efig
We may then ask how we are supposed to think about the orthogonal complement to the HaPPY code subspace: does it have a black hole interpretation?  Indeed it does: the idea is that we can start introducing black hole states by removing tensors.  For example consider the situation in figure \ref{tensorbh}, where we remove only the central tensor.  This corresponds to a black hole whose size is of order the AdS radius.  Away from the black hole bulk locality continues to function as usual, but near the black hole we have lost some of the spacetime.  As we make the black hole bigger by removing more tensors, it is straightforward to see that its coarse-grained entropy scales with its horizon area.\footnote{This is a bit too fast, since in hyperbolic space volume and area scale similarly.  The real point is that we are measuring the area in Planck units, as opposed to some combination of Planck and AdS units, which here corresponds to saying that the number of degrees of freedom per unit area on the black hole horizon is the same as the number of degrees of freedom at each boundary point (this is the $N^\sharp$ scaling of the previous subsection).  We will see below that the area relevant for the horizon is the same as that relevant for the Ryu-Takayanagi formula, which fixes the units.  This is more explicit in the random tensor models we discuss momentarily, since these have a large parameter we can make use of.}  Eventually we have removed all of the tensors, and are left with just the identity map on the boundary Hilbert space: we have thus eventually accounted for all of the states in the boundary theory.  Thus we see that the HaPPY code illustrates all of the general features we discussed in the previous subsection.  It does \textit{not} however have several other important features of holography: most importantly it does not have any natural boundary dynamics, and in the bulk the degrees of freedom are not local at distances which are small compared to the AdS radius.  For these reasons, it is important to remember that the HaPPY code is only an illustration of the connection between holography and quantum error correction, which as we have seen above rests on more general footing.

\bfig
\includegraphics[height=7cm]{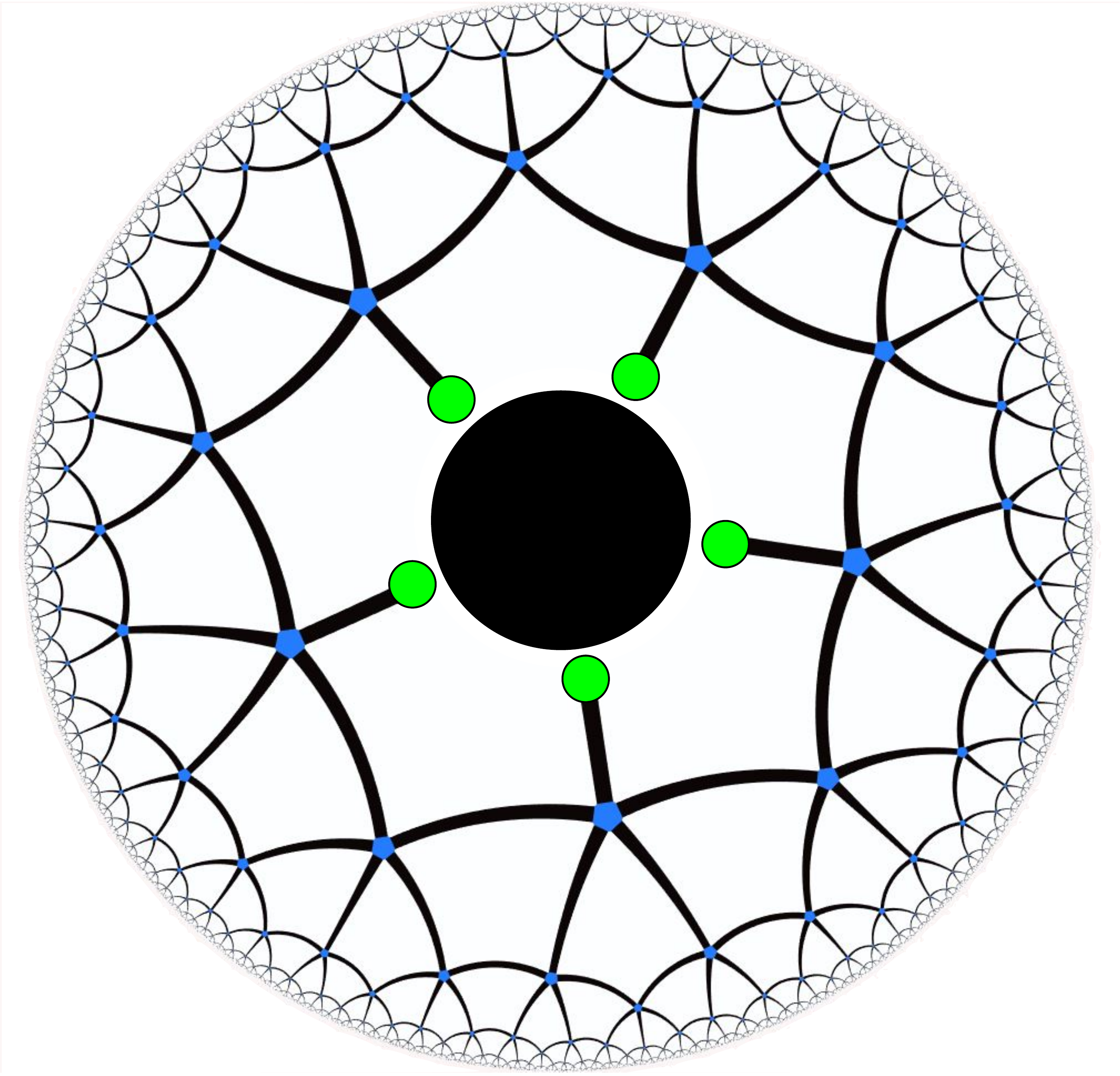}
\caption{A tensor network black hole.  One ``effective field theory'' index is replaced by five new ``black hole microstate'' indices, shaded green.  }\label{tensorbh}
\efig
Although lack of time prevents a substantial discussion, it is worth mentioning that there is a technical problem with the HaPPY code that can be avoided by replacing the perfect tensors with \textit{random} tensors, which are chosen independently at each vertex of the network \cite{Hayden:2016cfa}.  The problem is that, although the connection to subsystem quantum error correction with complementary recovery works for the choice of subregion shown in figure \ref{cutnetworkfig}, it does not quite work for all regions in the HaPPY code; there are choices of boundary region $R$ where there exist bulk degrees of freedom which can be recovered neither on $R$ or on $\ol{R}$ due to discrete artifacts in the network.  This problem goes away for the random tensor construction of \cite{Hayden:2016cfa}, where it was beautifully shown that in the limit where the range of the contracted indices is large, the network defines a subsystem quantum error correcting code with complementary recovery to high accuracy for any region $R$ and its complement.  

Finally I should mention that soon after \cite{Almheiri:2014lwa} came out, \cite{Mintun:2015qda} appeared, which used a simple $1+1$ dimensional model to argue that imposing gauge constraints in the boundary theory might lead to ``enough'' nonlocal entanglement to allow for the emergence of the radial direction.  This however is not true: roughly speaking since gauge constraints are local, at best they can can generate only ``short-range'' entanglement.  One illustration of this is that local constraints could not possibly be sufficient to explain how the union of the two connected components of the boundary region $R$ in figure \ref{twointervalsfig} below could have access to the shaded green region in the figure when the connected components separately can only access the red regions (see \cite{Pastawski:2015qua} for a quantitative explanation).  The right interpretation of the model in \cite{Mintun:2015qda} is that the reconstruction properties they found really followed from restricting to a code subspace of states whose energies are at most $O(N^0)$, and restricting to low energy states of a local scale invariant Hamiltonian \textit{does} create enough long-range entanglement to get an emergent radial direction.

\section{The Quantum Ryu-Takayanagi Formula}\label{rtsec}
In this section I'll discuss a subject which will at first seem unrelated to bulk reconstruction, the quantum Ryu-Takayanagi formula \cite{Ryu:2006bv,Hubeny:2007xt,Faulkner:2013ana}. We will soon see however that the connection between the two is quite close.  In fact we will see that the quantum Ryu-Takayanagi formula is essentially equivalent to the idea that a boundary subregion $R$ is dual to its \textit{entanglement wedge} $W[R]$ \cite{Harlow:2016vwg}, a bulk region which we will see contains the causal wedge $C[R]$ but is typically larger \cite{Wall:2012uf,Headrick:2014cta}.  Since there is an independent derivation of the quantum Ryu-Takayanagi formula \cite{Lewkowycz:2013nqa,Faulkner:2013ana,Dong:2013qoa,Dong:2016hjy}, this will establish subregion duality in the entanglement wedge \cite{Dong:2016eik,Harlow:2016vwg}. Along the way we will learn more about the meaning and range of validity of the RT formula.   

\subsection{RT Formula and the Entanglement Wedge}\label{RTsec}
Before stating the modern version of the Ryu-Takayanagi formula, we first need to introduce a few geometric concepts.

\bfig
\includegraphics[height=6cm]{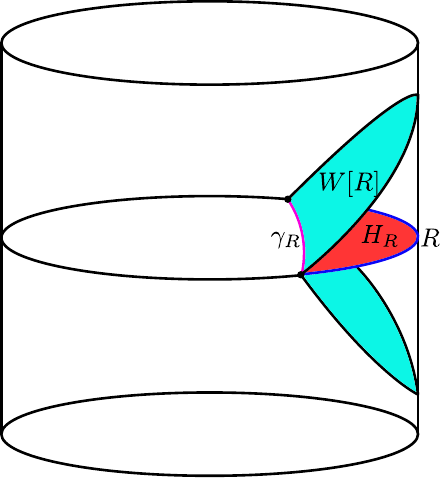}
\caption{The HRT surface $\gamma_R$ and entanglement wedge $W[R]$ of a boundary subregion $R$ for $AdS_3$.  Here $R$ extends between the two black dots, and is shaded blue.  $\gamma_R$ is shaded purple, a homology hypersurface $H_R$ is shaded red, and $W[R]$ is the spacetime volume between the two light blue surfaces.}\label{hrtfig}  
\efig
\begin{mydef}\label{HRTdef} Let $R$ be a boundary spatial subregion of some asymptotically-AdS geometry.   A \textbf{Hubeny-Rangamani-Takayanagi (HRT) surface for $R$} \cite{Hubeny:2007xt} is a codimension two bulk spatial submanifold with boundary, $\gamma_R$, with the following properties:
\bi
\item The boundary of $\gamma_R$ is equal to the boundary of $R$
\item The area of $\gamma_R$ is extremal under small variations of its location in spacetime, provided we restrict to variations which preserve $\partial \gamma_R=\partial R$
\item $\gamma_R$ is homologous to $R$, in the sense that there exists a spatial codimension-one submanifold with boundary, $H_R$, called a homology hypersurface, such that $\partial H_R=\gamma_R\cup R$
\item There is no other surface of strictly smaller area obeying the above three properties
\ei
\end{mydef}
There are a number of mathematical subtleties related to the existence and uniqueness of such surfaces, see \cite{Wall:2012uf,EngelHar}, but I will here just assume both.
\begin{mydef}
Let $R$ be a boundary spatial subregion of some asymptotically-AdS geometry. The \textbf{entanglement wedge of R}, $W[R]$, is defined as the bulk domain of dependence $D[H_R]$ of any homology hypersurface $H_R$ for the HRT surface $\gamma_R$.
\end{mydef}
I illustrate these definitions in figure \ref{hrtfig}.  
\bfig
\includegraphics[height=5.5cm]{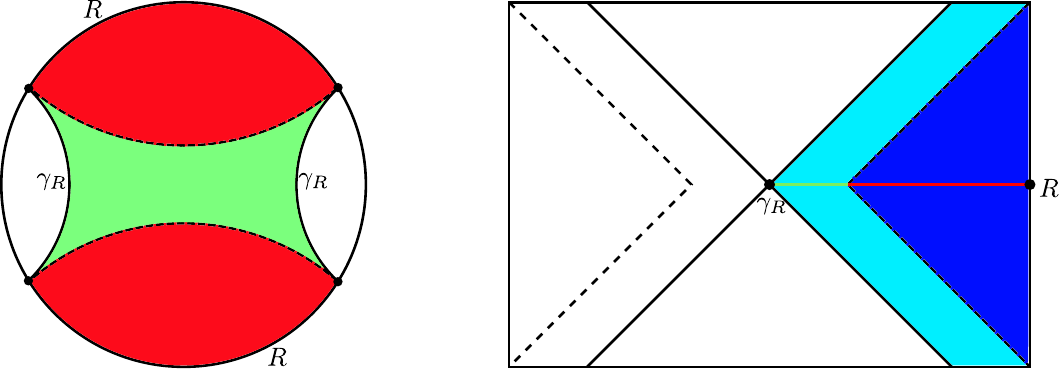}
\caption{Two examples of the entanglement wedge exceeding the causal wedge.  In the left diagram we have a time slice of empty $AdS_3$, and our boundary region $R$ is a union of two disjoint intervals which are sufficiently large that the solid lines are shorter than the dashed lines.  The homology hypersurface $H_R$ is the union of the red and green regions, while the causal wedge $C[R]$ contains only the red surfaces. The right diagram is a spacetime Penrose diagram for an ``elongated wormhole'', of the type that can be constructed using infalling matter as in \cite{Shenker:2013yza}.  $R$ is taken to be an entire timeslice of the right boundary, the causal wedge $C[R]$ is shaded dark blue, while the entanglement wedge $W[R]$ includes also the light blue points.  The homology hypersurface is again the union of the red and green points.  Note in particular that the entanglement wedge contains points behind the event horizon.}\label{twointervalsfig}
\efig
There is an immediate resemblance between the entanglement wedge and the AdS-Rindler wedge of figures \ref{rindlerfig}, \ref{multrindlerfig}.  This is not a coincidence: in empty AdS the entanglement wedge of a ball-shaped boundary region is equivalent to its AdS-Rindler wedge.  For more general regions and geometries the AdS-Rindler wedge does not make sense, but we can still discuss the causal wedge $C[R]$ from definition \eqref{causaldef}.  In fact there is a general theorem relating $C[R]$ and $W[R]$: under ``reasonable assumptions'' such as the null energy condition, global hyperbolicity, etc, we have the containment \cite{Wall:2012uf,Headrick:2014cta}
\be\label{CWcont}
C[R]\subset W[R].
\ee
In general the entanglement wedge can be quite a bit bigger than the causal wedge, I give two examples of this in figure \ref{twointervalsfig}.  An essential point is that the causal wedge of any boundary region can never contain points behind an event horizon, while the entanglement wedge can.  

We now have the tools to discuss the modern version of the Ryu-Takayanagi formula.  This formula was assembled gradually starting from the original proposal of \cite{Ryu:2006bv}, which applied only in static geometries and included no quantum effects.  This proposal was extended to dynamical spacetimes in \cite{Hubeny:2007xt}, which introduced the HRT surface, and then quantum effects were included finally in \cite{Faulkner:2013ana,Dong:2013qoa}.  The stationary classical version of the correspondence was given a reasonably precise derivation in \cite{Lewkowycz:2013nqa}, which was covariantized in \cite{Dong:2016hjy}, and which was extended to include quantum corrections in \cite{Faulkner:2013ana,Dong:2013qoa}.  I will simply define the Ryu-Takayanagi formula to be the outcome of this whole body of work: given a ``reasonable'' state $\rho$ in the boundary CFT, the von Neumann entropy of the reduced state $\rho_R$ obeys\footnote{This version of the formula is supposed to be correct up to terms that vanish with a positive power of $G$.  There is in fact a proposed generalization to all orders in $G$ \cite{Engelhardt:2014gca}, which has recently been argued for \cite{Dong:2017xht}, but there are many new subtleties which arise. Being correct at order $G^0$ will be good enough for us here.}
\be\label{RT}
S\left(\rho_A\right)=\tr\left(\rho \LR\right)+S\left(\rho_{H_R}\right),
\ee 
where 
\be
\LR\equiv \frac{\mathrm{Area}(\gamma_R)}{4G}+\ldots
\ee
is a bulk operator localized on $\gamma_R$ which starts out as the area of the HRT surface $\gamma_R$ in Planck units and then contains higher derivative corrections which we won't worry about \cite{Dong:2013qoa}.  $S(\rho_{H_R})$ denotes the bulk von Neumann entropy in the state $\rho$ of the fields on the homology hypersurface $H_R$.  We can also think of it more covariantly as the entropy of the state $\rho$ on the algebra of operators in the entanglement wedge $W[R]$.  

There are several mysterious features of this formula:
\bi 
\item What are the ``reasonable'' state supposed to be?  Can they include black holes?  Superpositions of geometries? The latter seems problematic because the first term on the right hand side is a linear functional of the state $\rho$, while the left hand side is a nonlinear functional.  We usually think of the first term as being dominant at small $G$, but how is this consistent with the linearity of quantum mechanics? \cite{Papadodimas:2015jra,Almheiri:2016blp}
\item Why should the area term be there at all?  If AdS/CFT is a duality with an isomorphism between states, then shouldn't entropy be dual to entropy? 
\item We've seen that the HRT surface $\gamma_R$ is generically outside of the causal wedge $C[R]$.  This does not seem to play nicely with our bulk reconstruction procedure based on solving bulk equations of motion, which didn't seem able to access bulk information outside of the causal wedge.
\ei
We will now see that all of these features are clarified by our quantum error correction picture of holography, and in particular we will see that the resolution of the third point is that for any CFT subregion $R$ we can in fact reconstruct the full algebra of operators in the entanglement wedge using CFT operators in $R$!  The possibility that the proper bulk subregion dual to a boundary subregion $R$ might be the entanglement wedge was suggested in \cite{Czech:2012bh,Wall:2012uf,Headrick:2014cta}, and the quantum error correction proof of this conjecture was given in \cite{Dong:2016eik}. Equation \eqref{CWcont} tells us that this idea is consistent with the bulk reconstruction we have been doing so far,  but the statement of entanglement wedge reconstruction tells us something much more interesting: at least in some cases there is no obstruction to reconstructing operators behind black hole horizons.

\subsection{RT formula in toy holography}\label{RTcode}
We'll begin by seeing how the RT formula arises in the three qutrit code.  Recall the encoding map \eqref{encode3}, which we can rewrite for logical mixed states as
\be
\wt{\rho}=U_{12}\Big(\rho_1 \otimes |\chi\ran\lan\chi|_{23}\Big)U_{12}^\dagger.
\ee
Here $\rho_1$ means we have written the logical mixed state onto the first physical qutrit. To test the RT formula, we should compute the von Neumann entropies of $\wt{\rho}$ reduced to various subsets of the physical qutrits.  This is easily done, by cyclic symmetry we just need to consider $\wt{\rho}_{12}$ and $\wt{\rho}_3$:
\begin{align}\nonumber
S\left(\wt{\rho}_3\right)&=\log 3\\
S\left(\wt{\rho}_{12}\right)&=\log 3+S\left(\wt{\rho}\right).\label{qutritRT}
\end{align}
If we define an ``area operator''
\be\label{LR1}
\LR\equiv \log 3,
\ee
then consulting figure \ref{qutritsfig1} we see that indeed the RT formula holds!  The logical degree of freedom in the center of the bulk is not in the ``entanglement wedge'' of the third physical qutrit, but it is in the ``entanglement wedge'' of the union of the first and second.  Therefore it should contribute bulk entropy only to $S(\wt{\rho}_{12})$, which is indeed what we see in \eqref{qutritRT}.

There are a few things to learn here.  First of all the RT formula is obeyed exactly, but only within the code subspace.  Within this subspace, there is no problem with linearity.  Secondly, we see that the area term arises from the entanglement in the state $|\chi\ran$: this entanglement is fixed in all code subspace states,  and we saw above that it is necessary for the redundancy of the code.  We thus see the resolution of the second puzzle described at the end of the previous subsection: the area term keeps track of the additional CFT entanglement which is not manifest in the low-energy bulk degrees of freedom.   

We can immediately uplift these observations to our tensor network models of holography using the machinery of subsystem quantum error correction with complementary recovery: consider figure \ref{cutnetworkfig}. The circuit diagram (or equivalently equation \eqref{compenc}) tells us that the von Neumann entropies of any state in the code subspace on the boundary regions $R$ and $\ol{R}$ are
\begin{align}\nonumber
S(\wt{\rho}_R)&=S(\chi_{R_2})+S(\wt{\rho}_r)\\
S(\wt{\rho}_{\ol{R}})&=S(\chi_{R_2})+S(\wt{\rho}_{\ol{r}}),
\end{align}
where $\chi_{R_2}\equiv\Tr_{\ol{R}_2}\left(|\chi\ran\lan\chi|\right)$.  Defining
\be\label{LR2}
\LR\equiv S(\chi_{R_2}),
\ee
we see that again we have a version of the RT formula, where the area term again represents the CFT entanglement not manifest in the low-energy bulk degrees of freedom.  Moreover we see that something quite beautiful happens: in either the HaPPY or the random tensor network, the state $|\chi\ran$ corresponds to a sequence of maximally entangled pairs on the links of the network cut by the green line $\gamma_R$.  So $S(\chi_{R_2})$ is just proportional to the number of cut links, or in other words the length of the line: it is the ``area operator'' indeed!  Moreover the constant of proportionality is just the logarithm of the Hilbert space dimensionality for each link: at the boundary we see that in the CFT this corresponds to the number of degrees of freedom at each lattice site.  But this is precisely the coefficient $N^\sharp$ appearing in \eqref{Geq}, \eqref{Sbulk}, so we see that we have indeed reproduced the correct holographic scaling of the entropy.

\subsection{Algebraic encoding and a theorem}
We've now seen several instances where quantum error correcting codes obey a version of the RT formula.  There is one feature of the coding RT formulas we've found so far which is somewhat unsatisfying however: in both cases the area operator is a $c$-number.  This is not consistent with our expectations from semiclassical gravity, where the area of the HRT surface is a nontrivial function on phase space.  This possibility can be accommodated into the formalism of quantum error correction, but we need to go beyond subsystem codes with complementary recovery.  In this final subsection I present without proof a set of mathematical results which illustrate the general relationship between subregion duality in the entanglement wedge and the RT formula: we will see that in fact they are precisely mathematically equivalent \cite{Harlow:2016vwg}.

The basic new ingredient we need has been mentioned in passing a few times already: we need to go from treating the bulk degrees of freedom in the entanglement wedge $W[R]$ as a tensor factor $\mathcal{H}_r$ of $\Hc$ to viewing them as a \textit{subalgebra} of the operators on $\Hc$.  The set of all operators on a tensor factor is one kind of subalgebra, but it is not the only kind.  The theory of operator algebras is in general quite intricate \cite{Jones}, but in finite-dimensional Hilbert spaces it is manageable.  Here I just give the relevant definitions and quote a few results, see e.g. the appendix of \cite{Harlow:2016vwg} for the proofs.\footnote{I have sometimes encountered resistance to treating von Neumann algebras on finite-dimensional Hilbert spaces as distinct from $C^*$ algebras.  This is because in finite dimension no discussion of topological closure is necessary, which to many is the real business end of von Neumann's definition.  But in fact the two are different even in finite dimension: a von Neumann algebra is \textit{defined} by its representation on a Hilbert space, and thus cannot have any inequivalent representations.  $C^*$ algebras are abstract objects, which can have inequivalent representations, and in particular the notion of the commutant, which is central to von Neumann algebras, has no counterpart for $C^*$ algebras.}
\begin{mydef}
Let $\Hh$ be a finite-dimensional complex Hilbert space, and $\mathcal{L}(\mathcal{H})$ be the set of linear operators on $\Hh$.  A subset $M\subset \Ll(\Hh)$ is a \textbf{von Neumann Algebra on $\Hh$} if the following hold:
\bi
\item For all $x$, $y$ in $M$, we have $xy\in M$
\item For all $x$, $y$ in $M$, we have $x+y\in M$
\item For all $x$ in $M$, we have $x^\dagger\in M$
\item For all $\lambda \in \mathbb{C}$, we have $\lambda I \in M$
\ei
\end{mydef}
In other words, a von Neumann algebra is is a subset of the operator algebra on $\Hh$ which is closed under addition, multiplication, and complex conjugation, and which contains all scalar multiples of the identity.  Any von Neumann algebra on $\Hh$ induces two other natural von Neumann algebras on $\Hh$:
\begin{mydef}
Let $M$ be a von Neumann algebra on $\Hh$.  Its \textbf{commutant}, $M'$, is the von Neumann algebra
\be
M'\equiv \{x\in \Ll(\Hh)|xy=yx \qquad\forall y \in M\}.
\ee
\end{mydef}
\begin{mydef}
Let $M$ be a von Neumann algebra on $\Hh$. Its \textbf{center}, $Z_M$, is the von Neumann algebra
\be
Z_M\equiv M\cap M'.
\ee
\end{mydef}
The commutant of $M$ is the set of all operators that commute with $M$, while the center is the subset of those which themselves are in $M$.  One simple example of these arises when the Hilbert space tensor factorizes as $\Hh=\Hh_r\otimes \Hh_{\ol{r}}$ and we take $M$ to be all the operators on $\Hh_r$.  We then have:
\begin{align}\nonumber
M&= \Ll(\Hh_r)\otimes I_{\ol{r}}\\\nonumber
M'&= I_r \otimes \Ll(\Hh_{\ol{r}})\\
Z_M&= \lambda I_{r\ol{r}}.
\end{align}
In this situation we say that $M$ is a \textit{factor}, and we see that any factor has a trivial center consisting of only scalar multiples of the identity.  In fact the converse is also true: any von Neumann algebra with trivial center is a factor, and having trivial center is actually usually taken as the definition of a factor.  Not all von Neumann algebras are factors, but the general case isn't so different:
\begin{thm}\label{vnthm}
Let $M$ be a von Neumann algebra on a finite-dimensional Hilbert space $\Hh$.  Then there exists a Hilbert space direct sum decomposition
\be\label{blockdecomp}
\Hh=\oplus_\alpha \Big(\Hh_{r_\alpha}\otimes \Hh_{\ol{r}_\alpha}\Big)
\ee
such that we have
\begin{align}\nonumber
M&=\oplus_\alpha\Big(\Ll(\Hh_{r_\alpha})\otimes I_{\ol{r}_\alpha}\Big)\\\nonumber
M'&= \oplus_\alpha\Big(I_{r_\alpha} \otimes \Ll(\Hh_{\ol{r}_\alpha})\Big)\\
Z_M&= \oplus_\alpha \Big(\lambda_\alpha I_{r_\alpha\ol{r}_\alpha}\Big).
\end{align}
\end{thm}
In other words, every von Neumann algebra on a finite-dimensional Hilbert space is a block diagonal direct sum of factors.  

The reader is surely already familiar with von Neumann's definition of the entropy of a state $\rho$ on a factor $M=\Ll(\Hh_r)\otimes I_{\ol{r}}$: we take the partial trace over $\Hh_{\ol{r}}$ and then compute
\be\label{factorent}
S(\rho,M)=-\Tr \left(\rho_r\log \rho_r\right)\equiv S(\rho_r).
\ee
This definition has a natural extension to an entropy of a state $\rho$ on a general von Neumann algebra $M$.  The basic idea is to consider the diagonal blocks of $\rho$ in the block decomposition \eqref{blockdecomp}:
\be
\rho=\begin{pmatrix}
p_1 \rho_{r_1\ol{r}_1} & \ldots & \ldots\\
\vdots & p_2 \rho_{r_2\ol{r}_2} & \ldots\\
\vdots & \vdots & \ddots
\end{pmatrix},
\ee
where we have extracted a positive coefficient $p_\alpha$ from each diagonal block so that $\Tr\rho_{r_\alpha \ol{r}_\alpha}=1$.  The $p_\alpha$ obey $\sum_\alpha p_\alpha=1$.  We then define the entropy of the state $\rho$ on $M$ as
\be
S(\rho,M)=-\sum_\alpha p_\alpha \log p_\alpha+\sum_\alpha p_\alpha S(\rho_{r_\alpha}),
\ee
where $\rho_{r_\alpha}=\Tr_{\ol{r}_\alpha}\rho_{r_\alpha \ol{r}_\alpha}$.  This entropy has two contributions: a ``classical'' piece associated to the uncertainty over which block we are in, and a ``quantum'' piece where we average the original von Neumann entropy \eqref{factorent} over blocks. This definition can be motivated using similar arguments to those which motivated \eqref{factorent}, and it obeys appropriate generalizations of the standard properties of von Neumann entropy (see the appendix of \cite{Harlow:2016vwg} for some discussion of this).  There is also an algebraic version of the \textit{relative entropy} of two states $\rho$ and $\sigma$ on an algebra $M$, which when $M$ is a factor is defined as
\be
S(\rho|\sigma,M)=\Tr\left(\rho_r\log \rho_r\right)-\Tr \left(\rho_r \log \sigma_r\right)\equiv S(\rho_r|\sigma_r).
\ee
Relative entropy is a measure of how much the states $\rho$ and $\sigma$ can be distinguished by measuring elements of $M$.  Its definition for an arbitrary von Neumann algebra $M$ is 
\be
S(\rho|\sigma,M)=\sum_\alpha p_\alpha^{\{\rho\}}\log \frac{p_\alpha^{\{\rho\}}}{p_\alpha^{\{\sigma\}}}+\sum_\alpha p_\alpha^{\{\rho\}}S\left(\rho_{r_\alpha}|\sigma_{r_\alpha}\right).
\ee

We are now in a situation where we can state the main theorem relating subregion duality in the entanglement wedge and the Ryu-Takayanagi formula \cite{Harlow:2016vwg}:
\begin{thm}\label{bigthm}
Let $\Hh$ be a finite dimensional Hilbert space which tensor factorizes into $\Hh_R\otimes \Hh_{\ol{R}}$, let $\Hc$ be a subspace of $\Hh$, and let $M$ be a von Neumann algebra on $\Hc$.  Then the following three statements are equivalent:\footnote{Note that this theorem does not promise that any of the three statements is true; it says only that they are either all true or all false.}
\bi
\item[(1)] For all operators $\wt{O}\in M$ and $\wt{O}'\in M'$, there are operators $O_R\in \Ll(\Hh_R)$ and $O_{\ol{R}}'\in \Ll(\Hh_{\ol{R}})$ such that for all states $|\wt{\psi}\ran\in \Hc$ we have
\begin{align}\nonumber
O_R|\wt{\psi}\ran&=\wt{O}|\wt{\psi}\ran\\\nonumber
O_R^\dagger|\wt{\psi}\ran&=\wt{O}^\dagger|\wt{\psi}\ran\\\nonumber
O_{\ol{R}}'|\wt{\psi}\ran&=\wt{O}'|\wt{\psi}\ran\\
O_{\ol{R}}'^\dagger|\wt{\psi}\ran&=\wt{O}'^\dagger|\wt{\psi}\ran
\end{align}
\item[(2)] There exists an operator $\LR\in Z_M$ such that for any state $\wt{\rho}$ on $\Hc$, we have
\begin{align}\nonumber
S(\wt{\rho}_R)&=\Tr\left(\wt{\rho}\LR\right)+S(\wt{\rho},M)\\
S(\wt{\rho}_{\ol{R}})&=\Tr\left(\wt{\rho}\LR\right)+S(\wt{\rho},M').\label{codeRT}
\end{align}
\item[(3)] For any states $\wt{\rho}$, $\wt{\sigma}$ on $\Hc$, we have
\begin{align}\nonumber
S(\wt{\rho}_R|\wt{\sigma}_R)&=S(\wt{\rho}|\wt{\sigma},M)\\
S(\wt{\rho}_{\ol{R}}|\wt{\sigma}_{\ol{R}})&=S(\wt{\rho}|\wt{\sigma},M')
\end{align}
\ei
\end{thm}
The application of this theorem to AdS/CFT works as follows: $\Hh$ is the full Hilbert space of the boundary CFT, and $\Hh_R$ and $\Hh_{\ol{R}}$ describe the CFT degrees of freedom in a boundary subregion $R$ and its complement $\ol{R}$.\footnote{Strictly speaking we need to regulate the CFT at some high energy scale to make $\Hh$ finite-dimensional, but this scale can be taken to be much higher than any other scale of interest for physics.}  $\Hc$ describes the code subspace on which some set of bulk observables we are interested in are expected to make sense, for example it could be the code subspace of the three qutrit code or the code subspace of the HaPPY code, or it could be the set of states in $\mathcal{N}=4$ super Yang Mills theory on spatial $\mathbb{S}^{3}$ which have energy at most $\lambda^{1/5}$.  $M$ describes the algebra of bulk operators in the entanglement wedge $W[R]$ of $R$, since this is the algebra for which the RT formula has been proven, and $M'$ describes the algebra of bulk operators in $W[\ol{R}]$.  Condition (1) is then the statement of subregion duality between $R$ and $W[R]$: it says that all entanglement wedge operators can be represented in $R$.  Condition (2) is the quantum Ryu-Takayanagi formula, with the added insight that the ``area operator'' $\LR$ must be in the center of the algebra $M$.  Condition (3) has also appeared in holography before, it is called the JLMS formula \cite{Jafferis:2015del}.  The theorem then says that all of these features of holography, each of which was understood somewhat independently, are in fact equivalent!  The proof of the theorem is fairly technical, so I will just refer to \cite{Harlow:2016vwg} for the details.  It is important to note however that several of the logical implications predate \cite{Harlow:2016vwg} in some form or other.  The implication $(1)\implies(2)$ was noticed in examples such as the  HaPPY and random tensor network codes already in \cite{Pastawski:2015qua,Hayden:2016cfa}, with earlier hints going back to \cite{Swingle:2009bg}. The implication $(2)\implies(3)$ was argued for somewhat heuristically in \cite{Jafferis:2015del}, and more rigorously for the case where $M$ is a factor in \cite{Dong:2016eik}, and the implication $(3)\implies(1)$ was also established, again for the special case where $M$ is a factor, in \cite{Dong:2016eik}.  The primary contributions of \cite{Harlow:2016vwg} were to generalize to arbitrary von Neumann algebras, to demonstrate that the implication $(1)\implies(2)$ was not tied to special examples, and to combine all these implications together in one package.

This theorem has many important consequences for holography, and I do not have time to really do any of them justice.  I will instead just sketch of a few, see \cite{Harlow:2016vwg} for more details.
\bi
\item Since the Ryu-Takayanagi formula has been independently established using the replica trick methodology of \cite{Lewkowycz:2013nqa,Dong:2016hjy,Faulkner:2013ana,Dong:2013qoa}, theorem \eqref{bigthm} establishes subregion duality in the entanglement wedge once and for all: we now know precisely which bulk subregion is dual to any boundary subregion.   So far this is probably the biggest achievement of the quantum error correction perspective on holography.
\item We see that to get a nontrivial area operator, meaning an area operator which is not proportional to the identity, it is essential that we take $M$ to be an algebra with nontrivial center.  In the bulk this is related to the fact that there are diffeomorphism constraints which present the Hilbert space from factorizing.  
\item The situation described in condition (1) is a generalization of the subsystem quantum correction with complementary recovery which I discussed around equation \eqref{compenc}: I call it \textit{operator algebra quantum error correction with complementary recovery}.  One can view theorem \eqref{bigthm} as the basic result characterizing this situation: in particular along the way to the proof, one finds the following generalization of equation \eqref{compenc}:
\be
|\wt{\alpha,ij}\ran=U_R U_{\ol{R}}\left(|\alpha,i\ran_{R_1^\alpha}|\alpha,j\ran_{\ol{R}_1^\alpha}|\chi_\alpha\ran_{R_2^\alpha\ol{R}_2^\alpha}\right),
\ee
in terms of which we can get a formula for the area operator $\LR$ analogous to equations \eqref{LR1}, \eqref{LR2}:
\be
\LR=\oplus_\alpha \left(S\left(\chi_{R_2^\alpha}\right) I_{r_\alpha \ol{r}_\alpha}\right).
\ee
Here comparing to theorem \eqref{vnthm} we see explicitly that $\LR$ is in the center of $M$.
\item We can use the framework of operator algebra quantum error correction with complementary recovery to study what happens if we include superpositions of geometries and/or black holes in the code subspace.  In such situations it is important to discuss properly-dressed diffeomorphism-invariant observables, but these are naturally accommodated in an algebraic framework.  In the former case we see that there is no problem with the RT formula continuing to hold in superpositions of geometries, which here roughly speaking correspond to states with projections onto more than one $\alpha$.  In particular the ``entropy of mixing'' terms which arise at $O(G^0)$ as pointed out in \cite{Papadodimas:2015jra,Almheiri:2016blp} are naturally reproduced by the ``classical'' piece of the algebraic entropy appearing in condition (2).  If we include black holes in the code subspace, there is an important decision about whether or not we include all of their distinct microstates, which has important consequences for how the ``homology condition''  in definition \eqref{HRTdef} should be applied; as long as we apply it correctly there does not seem to be any problem with the RT formula continuing to hold in such code subspaces`.  The quantum RT formula is really thus quite general, much more general than originally expected.  
\item Part of the choice of code subspace in the CFT is dual to the choice of short-distance cutoff in the bulk effective field theory.  For example in equation \eqref{codeRT}, the left hand side depends only on the state $\wt{\rho}$ and the region $R$, it does not depend on our choice of code subspace, while both terms on the right hand side \textit{do} depend on the choice of code subspace.  This is a coding analogue of the standard observation that entropy can be passed between the bulk and the area terms by changing the UV cutoff \cite{Solodukhin:2011gn}. 
\ei
Each of these ideas is still in development, and I believe they all would benefit from more work.  One avenue which has already seen important progress is formulating an ``approximate'' version of theorem \eqref{bigthm} \cite{Cotler:2017erl}.

\section{Conclusion}
It is my personal feeling that in the last few years our understanding of the emergence of the bulk in AdS/CFT has grown enormously.  The emergent radial direction has gone from a mystery to a well-understood phenomonenon, with simple toy models and powerful general results describing how it works in considerable detail.  Nonetheless there are still major gaps in our understanding of holography in AdS/CFT and beyond, so I will close by listing some which I feel are especially important.  
\bi
\item \textbf{Sub-AdS Locality:}  A general feature of exactly solvable models of holography so far, including not just the tensor network models described here but also higher spin \cite{Klebanov:2002ja} and $d=1$ \cite{Kitaev:2017awl,Maldacena:2016hyu} examples, is that they do not have local physics in the bulk at scales which are parametrically smaller than the AdS radius.  For tensor network models this problem arises because there is more to holography than spatial entanglement: at each boundary point there are $N^\sharp$ degrees of freedom, and sub-AdS locality puts strong constraints on how they interact with each other (see e.g. \cite{Bao:2015uaa} for some discussion of this).  For example in the $\mathcal{N}=4$ theory we only get sub-AdS locality in the bulk in the limit where the 't Hooft coupling $\lambda$ is very large, but none of the exactly soluble models have such a parameter.  It is not clear what the missing idea is to understand the structure of holography at sub-AdS scales, but one promising avenue is to study more deeply the emergence of the bulk in the BFSS matrix model which is dual to eleven dimensional flat Minkowski space \cite{Banks:1996vh}. It is an indication of how little we understand this problem that we do not even know whether or not supersymmetry, something I have neglected entirely in these notes, is a required feature (see \cite{Ooguri:2016pdq} for a recent claim that it is).
\item \textbf{Time Evolution:} In these notes we have basically always discussed the structure of holography at fixed time in the Schrodinger representation.  Boundary CFT dynamics have played little role in our discussion.  To some extent this is necessary: most reasonable choices for the code subspace are not preserved under time evolution.  In fact it is widely expected that in general relativity \textit{any} nontrivial classical initial data in AdS eventually leads to black hole collapse, which in the boundary CFT is dual to the statement that any perturbation should eventually thermalize.   At longer times all sorts of interesting things happen in the bulk which do not seem to be describable within bulk effective field theory: black holes return their information, long wormholes become short again, etc (see \cite{Cotler:2016fpe} for one recent discussion of these things).  Somehow the small errors in our CFT realization of the bulk accumulate as time evolves, and understanding how this happens is very important for understanding the resolution of the black hole information problem.  
\item \textbf{Gravitational Dressing:} In these notes I have mostly brushed off dealing with the issue of gravitational dressing of bulk local operators.  But actually this is very important in situations where backreaction is large.  These are precisely the situations where there is a distinction between the causal and entanglement wedges, and the recent exciting work on chaos and traversable wormholes in large part hinges on how bulk operators move around as we change the geometry \cite{Shenker:2013pqa,Gao:2016bin,Maldacena:2017axo}. I expect that gravitational dressing will be essential to understanding the proper description of the black hole interior.  Which leads us to:
\item \textbf{The Black Hole Interior:} The million dollar question in bulk reconstruction in AdS/CFT remains the description of the black hole interior in generic microstates \cite{Almheiri:2013hfa,Marolf:2013dba,Harlow:2014yka}.  Are there firewalls?  If so in which states?  Does quantum mechanics break down for the infalling observer?  We still don't know the answers to these questions, and until we do we certainly can't claim to have a complete theory of quantum gravity in asymptotically anti de Sitter space.  There is grounds for optimism, in particular the establishment of subregion duality in the entanglement wedge tells us that the horizon is not an impenetrable barrier for AdS/CFT, but much remains to be understood.  Solving all three of the previous problems will most likely be necessary to get a complete understanding.
\item \textbf{What about the real world:}  Some day we are going to have to confront the observational fact that the cosmological constant in our local region of the universe is not negative.  It is embarrassing that for twenty years so much effort has relied on ignoring this.  I console myself with the observation that there are many conceptual similarities between life in the black hole interior and life in de Sitter space, and that thus understanding the previous problem is a prerequisite for understanding this one, but there are also many differences.  Perhaps the key one is that in the most plausible theoretical cosmological scenario to date,\footnote{This statement is controversial in some quarters, but I think it is true \cite{Polchinski:2015pzt,Polchinski:2016xto,harlowinf}.} eternal inflation on the landscape of string theory, we are always right in the thick of things, with no simple reference to a stable place like the AdS boundary where things might be better-defined.  The best candidate for such a place seems to be supersymmetric terminal vacua with zero cosmological constant \cite{Susskind:2007pv,Harlow:2010my,Harlow:2012dd}, but we are very far away from understanding or using such a description. Perhaps a better strategy would be to learn how to talk with some imprecision about observables directly within the bulk, and indeed this is the strategy we have been pursuing in these lectures in the relatively sheltered context of AdS/CFT.

\ei

\paragraph{Acknowledgments} I first would like to thank the organizers Mirjam Cvetic and Igor Klebanov for the opportunity to speak at TASI 2017, as well as the local organizers Tom DeGrand and Oliver DeWolfe for running a remarkably slick operation.  I have very fond memories of my own time as a student at TASI 2010, and it was a pleasure to see things from the other side.  I'd like to thank also the students and the other lecturers, with whom I had many interesting discussions, as well as fun social outings out and around Boulder.  Earlier versions of these lectures were given at the Asian Winter School at OIST in 2016 and at TIFR Mumbai in 2017, and they were presented once again at the PiTP summer school at IAS in 2018, so I thank those institutions for excellent hospitality as well.  I owe a lot of what I know about the topics discussed here to many people, mentioning them all would be impossible, but certainly I must acknowledge Ahmed Almheiri, Raphael Bousso, Bartek Czech, Xi Dong, Netta Engelhardt, Patrick Hayden, Matt Headrick, Daniel Jafferis, Dan Kabat, Hong Liu, Juan Maldacena, Don Marolf, Hirosi Ooguri, Fernando Pastawski, Eric Perlmutter, Joe Polchinski, John Preskill, Andrea Puhm,  Steve Shenker, Eva Silverstein, Douglas Stanford, Andy Strominger, Lenny Susskind, Herman Verlinde, Aron Wall, Edward Witten, and Beni Yoshida. I'd also like to thank Geoff Penington for a discussion which led to a considerable improvement of section 4.3.  My work is supported by the Simons Foundation through the ``It from Qubit Collaboration'', as well as DOE grant DE-SC0007870 (through last June) and the MIT physics department. 

Finally I dedicate these notes to the memory of Joe Polchinski.  Losing him so early is a tragedy in the truest sense of the word.  Joe was a constant inspiration for me over the last ten years, from my student days reading his textbook and many classic papers, to listening to his lectures at TASI 2010, to our early discussions about firewalls (he was right), to our recent battle over the causal wedge vs. the entanglement wedge and boundary gauge symmetry in holography (I was right), to him smoking me up and down Old San Marcos in Goleta on his bike.  His influence appears on every page of these notes, which describe a subject which was dear to him for twenty years.  Often when writing a paper, I would think of Joe as my target audience: he didn't want to see unnecessary details, but he didn't fall for propaganda either.  His recent memoirs are a must-read for anyone who cares about theoretical physics \cite{Polchinski:2017vik}. Joe was truly one of the greats in the history of science, and his contributions will be felt far into the future.  But more than that, he was an outstanding human being: his clarity of thought was matched only by his humility and his mischievous sense of fun. We are all the poorer in his absence.   

\bibliographystyle{jhep}
\bibliography{bibliography}
\end{document}